\documentclass[prx, 10pt, twocolumn, floatfix, superscriptaddress] {revtex4-1}
\usepackage{times, mathrsfs, amsmath, amsfonts, graphics, graphicx, cancel, color, xcolor, theorem, bbm, mathtools, amssymb, xargs, blkarray}
\usepackage{kbordermatrix}
\usepackage{multirow, array}
\usepackage[english]{babel}
\usepackage[margin=.75in]{geometry}
\usepackage[T1]{fontenc}
\usepackage{xfrac,relsize,float}
\usepackage[unicode=true, breaklinks=false, pdfborder={0 0 1}, backref=false, colorlinks=true, linkcolor=blue, citecolor=blue]{hyperref}
\usepackage{algorithm}
\usepackage{algorithmic}
\usepackage{siunitx}
\usepackage{amsmath,stackrel}

\newcommand{\eM}{\ensuremath{\epsilon\text{-machine}}}
\newcommand{\eMs}{\ensuremath{\epsilon\text{-machines}}}
\renewcommand{\arraystretch}{1.5}

\renewcommand{\v}[1]{\ensuremath{\mathbf{#1}}} 
 
\newcommand{\cev}[1]{\reflectbox{\ensuremath{\vec{\reflectbox{\ensuremath{#1}}}}}} 
\renewcommand{\kbldelim}{(}
\renewcommand{\kbrdelim}{)}

\begin{document}

\title{Memory in quantum dot blinking}

\author{Roberto N. Mu{\~n}oz}
\email{roberto.munoz@monash.edu}
\affiliation{ARC Centre of Excellence in Exciton Science and School of Physics \& Astronomy, Monash University, Clayton, Victoria 3800, Australia}

\author{Laszlo Frazer}
\affiliation{ARC Centre of Excellence in Exciton Science and School of Chemistry, Monash University, Clayton, VIC 3800, Australia}

\author{Gangcheng Yuan}
\affiliation{ARC Centre of Excellence in Exciton Science and School of Chemistry, Monash University, Clayton, VIC 3800, Australia}

\author{Paul Mulvaney}
\affiliation{ARC Centre of Excellence in Exciton Science, School of Chemistry, University of Melbourne, Australia}

\author{Felix A. Pollock}
\affiliation{School of Physics \& Astronomy, Monash University, Clayton, Victoria 3800, Australia}

\author{Kavan Modi}
\affiliation{ARC Centre of Excellence in Exciton Science and School of Physics \& Astronomy, Monash University, Clayton, Victoria 3800, Australia}

\date{\today}
\begin{abstract} 
The photoluminescence intermittency (blinking) of quantum dots is interesting because it is an easily-measured quantum process whose transition statistics cannot be explained by Fermi's Golden Rule.  Commonly, the transition statistics are power-law distributed, implying that quantum dots possess at least trivial memories.  By investigating the temporal correlations in the blinking data, we demonstrate with high statistical confidence that quantum dot blinking data has non-trivial memory, which we define to be statistical complexity greater than one.  We show that this memory cannot be discovered using the transition distribution. We show by simulation that this memory does not arise from standard data manipulations.  Finally, we conclude that at least three physical mechanisms can explain the measured non-trivial memory: 1) Storage of state information in the chemical structure of a quantum dot; 2)  The existence of more than two intensity levels in a quantum dot; and 3) The overlap in the intensity distributions of the quantum dot states, which arises from fundamental photon statistics.

\end{abstract}

\maketitle

\section{Introduction} 
\label{sec:Introduction}

Quantum dots (\textbf{QDs}) are nanoscale semiconducting crystals~\cite{koch1993semiconductor,jacak2013quantum,bacon2014graphene} with a major role in developing energy-efficient technology~\cite{kouhnavard2014review,jun2013quantum}. When used as light-emitting diodes, QDs can be synthesized to emit light at precise colors which has resulted in the production of high-quality displays~\cite{moon2019stability}. An attractive advantage in using QDs for every-day and commercial use, is that they outperform conventional LEDs in terms of output and cost efficiency~\cite{wierer2016}, allowing for lighting spaces for less.

Despite these promises, the integration of QDs into scalable lighting applications is limited by intermittent photoluminescence known as blinking~\cite{Yuan2018,Kuno2000,nirmal1996fluorescence,lee2009brightening}. The blinking phenomenon is observed over a variety of QDs with different compositions~\cite{Frantsuzov2008,Vosch2001}, and results in periods where an excited QD enters an \emph{Off} state and no longer emits photons~\cite{Efros2016}. The intermittent lack of photons makes lighting less efficient~\cite{FerneeCSR2014,Frantsuzov2008}, and hence mitigating blinking is widely studied.

In order to mitigate blinking, it is necessary to understand the physics \& chemistry behind the phenomena. Blinking is generally characterised from data by employing statistical methods like duration distributions~\cite{Kuno2000,Plakhotnik2010}, power spectra~\cite{Pelton2004}, or autocorrelations~\cite{hou2020memories,Houel2015,stefani2005memory}, and a number of chemical and physical pathways have been proposed to contribute to blinking using these methods~\cite{brawand2015,Yuan2018,Efros2016,Vosch2001,Gomez2006,Kilina2016}. A possible way to better understand the blinking process is to look for correlations in the blinking patterns, i.e., characterise the blinking using information linking the past and future blinking patterns in a time series. Such correlations are often referred to as \emph{memory}~\cite{van1992stochastic}. Information theory reveals that the memory in a data sequence requires special tools to infer it, and is inaccessible via the aforementioned common statistical methods~\cite{riechers2021}. 

In this paper, we show that the blinking patterns of QDs indeed have memory. We do this by using methods from computational mechanics which infer memory from data~\cite{CrutchPRL1989, CSSR2, Ephraim2002}. We use hypothesis testing to show a model with trivial memory is a significantly worse explanation of the data than a model with nontrivial memory. Importantly, the model with trivial memory is rejected, even though it is an exact fit to the distribution of blinks. 

We then relate the observation of blinking memory to a number of processes which may be acting within the QDs in our analysis. Our methods are applicable to a wide range of chemical systems, and function as a flexible and scalable stepping point for future memory-based analyses. We propose that the ability for memory to be associated with specific chemical processes can ultimately serve as a diagnostic tool for the improvement of QD synthesis techniques, and overall output quality.

The manuscript is structured as follows: Section~\ref{sec:dataDesc} is a brief description of the QD blinking data used in this study, in addition to the elementary processing needed to separate the brightness states. Section~\ref{sec:powerlaws} covers the On/Off state durations and the statistics which they exhibit. It focuses on the power-law behaviour observed in some blinking QDs and how it relates to searches of memory. We then look at how memory is described in Section~\ref{sec:memoryIntro} and how it can be inferred from blinking data in Section~\ref{sec:HMMandCSSR}. We also cover how hypothesis testing can reject different models of memory in Section~\ref{sec:HypTesting}, before applying this methodology to the QD blinking data. The main results are shown in Section~\ref{sec:CGandResults}, followed by a discussion on potential sources for the observed memory in Section~\ref{sec:Discussion}, with simulated examples.

\section{Quantum dot data and treatments}

\begin{figure*}[th]
    \centering
    \includegraphics[width=\textwidth]
    {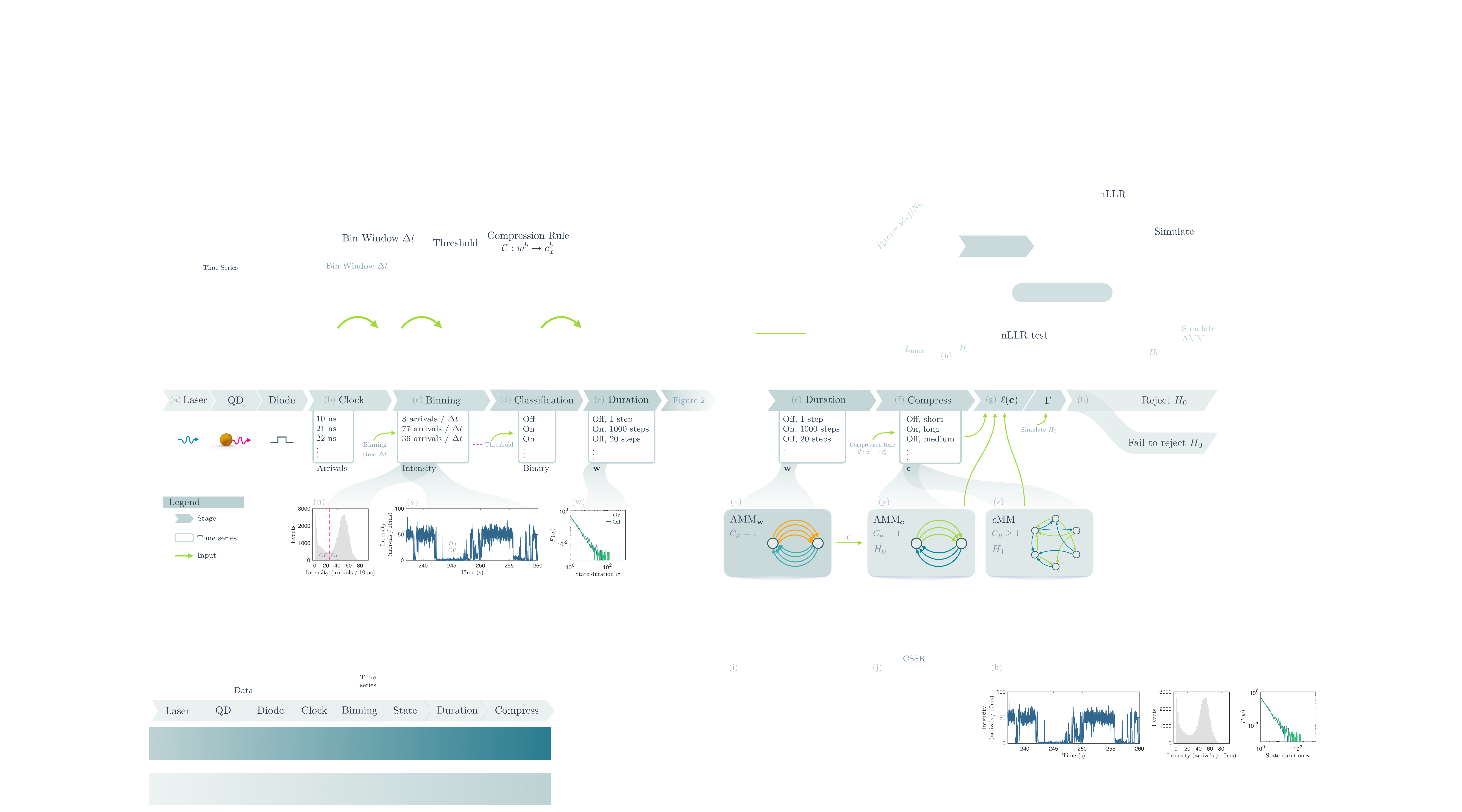}
    \caption{A flow chart representing data collection and processing. \textbf{(a)} A laser provides a beam of light which excites quantum dots that emit luminescence. The luminescence photons are converted to electrical pulses using an avalanche photodiode. 
    \textbf{(b)} A clock records the time when electrical pulses are received. 
    \textbf{(c)} The data is divided into time bins with width $\Delta t$, chosen to maximise contrast between the peaks in the histogram of intensities (u). The number of photon arrival times in each bin is computed. 
    \textbf{(d)} Each time bin is classified by using an intensity threshold $I_{\text{thresh}}$ to determine if the quantum dot is on or off during the time bin. 
    \textbf{(e)} The duration of on and off states is computed. 
    \textbf{(u-w)} Summary data taken from QD4@\SI{200}{\nano\watt}. (u) Intensity histogram obtained by binning arrival times over \SI{10}{\milli\second} windows. Red dashed line marks the intensity threshold, taken to be at the minima between the On and Off peaks. (v) Portion of the intensity trace of the blinking QD. Blue data indicates the recorded intensity over time with clear indications of the blinking intermittency. (w) Distribution of state durations graphed on a log-log scale. Data exhibits linear behaviour characteristic of power-law-like statistics.}
    \label{fig:workflow}
\end{figure*}

\subsection{Data collection and processing}
\label{sec:dataDesc}
We start by covering only the essential details of QD synthesis and measurement necessary for this study. Further elaboration may be consulted from Ref.~\cite{Yuan2018}.
Our study used core-shell CdSe/Cd$_{x}$Zn$_{1-x}$S quantum dots, previously synthesised~\cite{boldt2013synthesis,yuan2018tuning} and measured in Ref.~\cite{Yuan2018}. A total of 7 identically prepared quantum dots were each excited with a pulsed laser diode, and measured over excitation powers ranging from \SI{100}{\nano\watt} to \SI{600}{\nano\watt}. The photoluminescence of single QDs was collected using a microscope onto an avalanche photodiode detector. A nanosecond-resolution clock was then used to record the times of photon arrivals coming from an excited QD. We binned the photon arrival time measurements into 10ms intervals to convert them into a sequence of intensities (Figure~\ref{fig:workflow}c). An example of the intensity histogram for a blinking quantum dot is shown in Figure~\ref{fig:workflow}u. 

Most blinking dots generally feature two prominent peaks in intensity distributions: one corresponding to a \emph{bright}/\emph{On} state and the other corresponding to a \emph{dark}/\emph{Off} state. In the bright state, photons arrive in the detector with a rate $\lambda_{\text{On}}$, while the dark state has an arrival rate $\lambda_{\text{Off}}$ that is governed by background. Quantum dots with high emitting rates that spend the majority of the measured time in the On state typically exhibit a large amount of bright counts relative to dark. This is observed by the relative heights of the two peaks in the distribution, in addition to high contrast between the peaks. We found binning photon arrival times over 10ms windows to give the best contrast between intensity peaks. When investigating the nature of quantum dot blinking, being able to classify a series of intensity data into bright and dark states is a crucial step for understanding the underlying chemical processes. For each quantum dot, this was performed by selecting an intensity threshold $I_{\text{thresh}}$ located at the minima between the bright and dark intensity peaks. For each intensity outcome $I_k \in \v{I} = (I_1, I_2, \ldots, I_k,\ldots)$, we classified them into On/Off states using the threshold, $I_{\text{Off}} \leq I_{\text{thresh}} < I_{\text{On}}$, to form a sequence of classifications (Figure~\ref{fig:workflow}d).

\subsection{State durations and power-laws}
\label{sec:powerlaws}
The separation of intensity data into binary On/Off states is a common way to explore blinking quantum dots. For example, by scanning through the classification sequence and recording the length of consecutive On and Off outcomes, one obtains an alternating sequence of On/Off state durations 
\begin{gather}
    \mathbf{w} = (w_1^{\text{On}}, w_2^{\text{Off}}, \ldots, w_k^{b}, \ldots) \;:\; b\in\{\text{On},\text{Off}\},
    \label{eq:durationTimeseries}
\end{gather}
 shown in Figure~\ref{fig:workflow}e.
State durations, where the amount of time $w^b$ a quantum dot can spends in brightness state $b$ before blinking can be modelled as a pair of probability distributions $P_{b}(w)$. These distributions were regarded to be power-law distributed~\cite{Plakhotnik2010,Kuno2000,FerneeCSR2014,Frantsuzov2008,FerneeCSR2014}
\begin{gather}
    P_b(w^b) \propto (w^b)^{-m_{b}},
    \label{eq:powerlaw-general}
\end{gather}
where $m_{b}$ is the power-law constant, identified by linear behaviour when plotted on a log-log scale (Figure. 1). The most remarkable feature regarding power-law state durations and blinking behaviour, is the range in state durations over orders of magnitude~\cite{Lutz2004}. This means that quantum dots have been observed to blink at timescales orders of magnitudes in difference, from milliseconds to hours~\cite{Plakhotnik2010,sher2008}. Recent studies over a wider range of QD compositions and synthesis techniques have demonstrated the existence of exponential, multi-exponential, and quasi-exponential distributions for duration statistics~\cite{Yuan2018,ghosh2021}. Identifying the type of distribution present in blinking data has allowed experimenters to relate them to underlying chemical processes, e.g. exponential $P_{b}(w^b) \propto \exp (-C w)$ and power-law durations related to two different blinking mechanisms in Ref.~\cite{Yuan2018}.

Fermi's Golden Rule says that transitions between eigenstates of the Hamiltonian are exponentially distributed. This suggests that, for blinks which are not exponentially distributed, the on and off states are actually collections of multiple eigenstates. If those eigenstates correspond to different chemical structures, they may tend to occur in certain sequences.  For example, changes to surface bonding can enable changes one layer beneath the surface~\cite{stelmakh2017effect,meena2013understanding,fan2013sphere} Therefore, there is reason to suspect that additional information can be inferred from the blinking sequence $\v{w}$, than just the distributions from Equation~\eqref{eq:powerlaw-general} alone. 

Notably, the fact that state duration distributions can often be described by power-law statistics is intriguing in its own right. For example, in statistical mechanics, power-law distributions arises for systems at critical points such as during phase transitions~\cite{graves2017brief}. The appearance of power-law statistics in other fields is similarly attributed to other systems of interest whose dynamics are suspected to reside at criticality~\cite{landau1980,domb1996critical}, or that are scale invariant~\cite{stanley1999scaling}. Another central idea surrounding systems with power-law statistics is that they possess high degrees of \emph{memory}~\cite{Marzen2016}. That is, systems which depend on correlations over multiple points of time in the past~\cite{van1992stochastic}. While we unpack the notion of memory in the next section, the main takeaway here is that memory arises in a multitude of complex systems over a breadth of research fields. Some examples range from open quantum systems~\cite{pollock2018non,gennaro2009relaxation} to the firing patterns of neurons in the brain~\cite{Stringer2019,Munoz2020}. Finding equations which predict the future evolution of systems with memory is difficult, since the past behaviour of the system must be taken into account~\cite{milz2017introduction,breuer2002theory}. 

If observing power-law distributions alone truly is the ``litmus test" for memory, then one would expect there to be multi-time correlations in the power-law distributed QD blink durations. Ultimately, histogramming any temporal data on a log-log scale and observing a linear trend would be sufficient to mark the presence of memory, and no other additional tools would be needed. However, this is not the case. There is a body of research which counters the idea that power-law distributions imply a deep structure and high memory. Power-law distributions can arise from an ensemble of independent, random and unstructured processes, none of which at the individual level contain multi-time correlations~\cite{touboul2017power}. Notable examples include combining multiple exponential distributions together~\cite{reed2002gene,li1992random}, and even an ensemble of monkeys on typewriters~\cite{miller1957some}. Even using other statistical treatments such as autocorrelation~\cite{hou2020memories,Houel2015,stefani2005memory} or power spectra~\cite{Pelton2004} are shown to be insufficient to reveal memory in some cases~\cite{riechers2021}. 

Ultimately, the ability to infer the presence of memory in the state durations of blinking quantum dots depends on the use of specialised tools. In the following sections, we give an overview of the formalism needed to model memory in data, in addition to the tools needed to infer it. Applied to the sequence of state durations $\v{w}$ of real quantum dots, we examine whether or not quantum dots contain memory, and what kind of statements they allow one to make about the underlying blinking processes.

\section{Non-Markovianity, Computational Mechanics}
\label{sec:CompMech}
We aim to infer the memory present in the blinking quantum dot data that is inaccessible via state durations distributions alone. We describe the specialised tools necessary for this task, in addition to the mathematical framework they entail. We start by defining memory as the amount of past blinking information required to predict the future behaviour of the data (\textbf{non-Markovianity})~\cite{CrutchPRL1989}, and how it relates to the length of past observation sequences (Markov order)~\cite{breuer2002theory}. We then describe how models can efficiently characterise memory. Next, we explain how to construct these models from data, and which model corresponds to the memoryless data hypothesis. We then go on to outline the hypothesis testing methods used to reject the null model. Finally, we find that given the data, we can reject particular memoryless models compared to ones which include memory in the data.

\subsection{Time series and prediction}
\label{sec:memoryIntro}
Regardless of the physical system, inferring the memory of a process requires measurement data to be described as a \emph{time series}~\cite{Rabiner1989}. This means that at each time step $k$ the system is measured and the measurement outcomes $r_{k}$ at each step form a sequence $\mathbf{r} = (r_1, \ldots, r_{k}, \ldots)$. For the case of a blinking quantum dot, this could be the sequence of intensity measurements over time bins (Figure~\ref{fig:workflow}c), or the sequence of brightness state durations (Figure~\ref{fig:workflow}e). Each element of the data $\mathbf{r}$ is drawn from the set of measurement outcomes, or \emph{alphabet} $\mathcal{A}$, such that $r_k \in \mathcal{A}$. For most applications, the time series $\mathbf{r}$ is considered over discrete time steps rather than continuous, in addition to considering the set of measurement outcomes $\mathcal{A}$ to be finite. When the occurrence of a measurement outcome at a given time step is random, the system generating the time series is a stochastic process~\cite{DoobStochastic}. The sequential measurement of state durations $\v{w}$ mentioned in Equation~\eqref{eq:durationTimeseries} are an example of a stochastic process as they are drawn at random from power-law distributions following Equation~\eqref{eq:powerlaw-general}. It is important to note however, that although stochastic, measurement outcomes may not appear in a completely independent manner. The probability of seeing a particular outcome in the future may strongly depend on the sequence of outcomes observed in the past. This concept defines the notion of \emph{memory} when referring to stochastic time series. Being explicitly linked to past/future correlations, memory plays a central role in the prediction of the behaviour of a given stochastic process~\cite{van1992stochastic,breuer2002theory}.

Relative to an arbitrary time step $k$, we denote the future and the past of a time series as $\mathbf{r} = (\cev{r}, \vec{r})$. Explicitly the past and the future are $\cev{r} = (\ldots, r_{k-2},r_{k-1})$ and $\vec{r} = (r_{k}, r_{k+1}, \ldots)$ respectively. Processes which are memoryless follow the \emph{Markov property} 
\begin{gather}
    P(r_k \;|\; \cev{r}) = P(r_k \;|\; r_{k-1}),
    \label{eq:MarkovProperty}
\end{gather}
which states that future evolution depends solely on the current outcome and not on any of the past observations~\cite{van1992stochastic}. In general however, prediction of the immediate future outcome $r_k$, requires knowledge of the past $L$ outcomes $\cev{r}_{L} :=(r_{k-L}, \ldots, r_{k-2}, r_{k-1})$. Systems where this is true have memory, and are known as \emph{non-Markovian}. Formally stated, non-Markovian systems satisfy
\begin{gather}
    P(r_k \;|\; \cev{r}_L) \neq P(r_k \;|\; r_{k-1}),
    \label{eq:nonMarkovProperty}
\end{gather}
where the number of past outcomes $L$ needed to account for in order to optimally predict the future sequence is called the Markov order~\cite{Gagniuc2017}. For a sequence of state duration data $\v{w}$, this means that predicting the amount of time a quantum dot remains in either the On/Off state before blinking is dependent on how long it spent in the On/Off state at multiple instances in the past. Processes which follow Equation~\eqref{eq:MarkovProperty} are Markov processes and have $L=1$, which we will show is related to a memoryless hypothesis for the blinking.

\subsection{Memory modelled by HMMs}
\label{sec:HMMandCSSR}
We aim to perform hypothesis testing in order to reject a memoryless model in favour of a non-Markovian one. To do this, we need two models which represent the Markovian and non-Markovian hypotheses. In this section we outline how to construct these models which represent memory in the data.

\begin{figure*}[ht]
    \centering
    \includegraphics[width=0.85\textwidth]{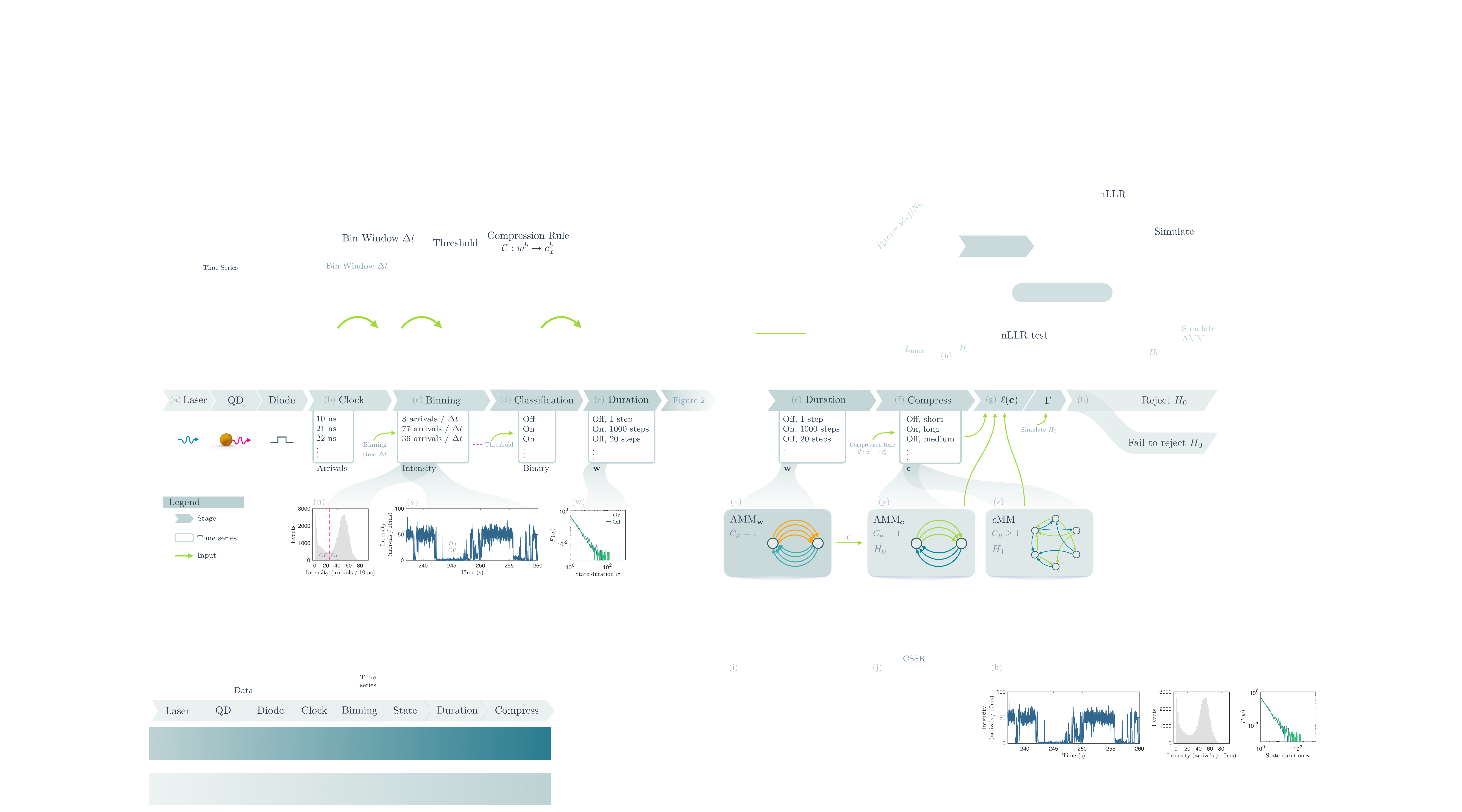}
    \caption{Flow chart representing the data compression, construction of the AMM and $\epsilon$MM models, and hypothesis testing. 
    \textbf{(e)} The duration of On and Off states is converted to a sequence of reduced measurement outcomes via the compression rule $\mathcal{C}$ in Equation~\eqref{eq:compressionRule} to meet the constraints described in Section~\ref{sec:HMMandCSSR}. 
    \textbf{(x)} The Alternating Markov Model for a time series of On/Off state durations $\v{w}$. Each gold arrow is associated with the probability for a specific On duration to be observed at the immediate future time step in the sequence. Blue arrows represent the same for an Off duration. The grey nodes are the causal states.
    \textbf{(f)} Time series of compressed state durations, ready for the hypothesis testing stage. 
    \textbf{(y)} The AMM fit to compressed data $\v{c}$. Each green arrow represents the probability for the coarse On duration to be observed at a future time step in the sequence. Blue arrows are the same for a coarse Off duration. This model is the result of performing the compression $\mathcal{C}$, and results in the same alternating structure, but with fewer measurement outcomes (arrows) as desired.
    \textbf{(z)} An example of an $\epsilon$MM inferred from the compressed data by the CSSR algorithm up to a Markov order $L_{\max}$. The graph structure shown here is an example of what is representative in QDs with memory. In general $\epsilon$MMs inferred from non-Markovian data have more nodes and are hence more complex than the AMM.
    \textbf{(g)} Hypothesis testing stage consisting of the negative log-likelihood $\ell(\v{c})$ and AMM rejection threshold $\Gamma$. The $\ell(\v{c})$ takes the compressed data $\v{c}$, AMM, and $\epsilon$MM models as inputs. The threshold is computed by simulating the AMM. 
    \textbf{(h)} Hypothesis test decision to either reject the null or fail to reject the null. For each time series data, $\ell(\v{c})$ is compared to the rejection threshold $\Gamma$ as shown in Equation~\eqref{eq:nLLRt}. Rejection of the null indicates significant non-Markovian correlations in the blinking data.}
    \label{fig:workflowHMM}
\end{figure*}

For non-Markovian systems, Equation~\eqref{eq:nonMarkovProperty} suggests that memory is related to the prediction of a process's future behaviour based off past observations, and is proportional to Markov order. Although the difficulty in predicting the future behaviour of non-Markovian processes rises exponentially with Markov order, there are efficient ways construct models that represent the underlying memory. For example, not all past observation sequences may result in distinct future behaviours, and hence not every possible past sequence must be remembered by a non-Markovian process. 

This was the central idea in work by Crutchfield and Young~\cite{CrutchPRL1989,epsilonMachines2}, whereby distinct past sequences that caused statistically equivalent future behaviour were grouped together. Formally, two distinct sequences of past observations $\cev{r}_{L}$ and $\cev{r}_{L}'$ of length $L$ belong to the same grouping $S_i \in \mathcal{S}$, if the probability of observing a specific future $\vec{r}$ given $\cev{r}_{L}$ or $\cev{r}_{L}'$ is the \textit{same}; that is 
\begin{gather}
     \cev{r}_{L} \sim_\epsilon \cev{r}_{L}' \quad\text{if}\quad P(\vec{r} \;|\; \cev{r}_{L}) = P(\vec{r} \;|\; \cev{r}_{L}'),
     \label{eq:equivRelation}
\end{gather}
where $\sim_\epsilon$ indicates that two past sequences correspond to the same partition. The entire collection $\mathcal{S}$ of all the partitions $S_i$ are known as the \emph{causal states} of the process. 

Due to the nature of their construction, new causal states are only created when necessary and hence relate to \emph{minimal} amount of memory a non-Markovian process needs to store to determine future behaviour. In the context of memory, the causal states are the hidden states that a non-Markovian system can be in.

The Shannon entropy~\cite{Shannon1948,Yeung2012} over the causal states quantifies the minimal number of bits of information required to optimally predict the future behaviour of the process. The measure introduced in Ref.~\cite{CrutchPRL1989}, is called the \emph{statistical complexity}:
\begin{gather}
    C_{\mu} := H[\mathcal{S}] = -\sum_i P(S_i) \log P(S_i),
    \label{eq:Cmu}
\end{gather}
and is the memory of a non-Markovian process.

When the causal states $\mathcal{S}$ are known, the probabilities to transition from one causal state to another $\mathcal{T}$ can be estimated from the data. A collection of states and transitions is a \emph{Hidden Markov Model} (\textbf{HMM})~\cite{DoobStochastic,Yeung2012} and allows the memory of a non-Markovian process to be represented by a digraph~\cite{CrutchPRL1989, Gagniuc2017}. The graph $G(V,E)$ consists of a set of vertices $v_i \in V$ and directed edges $\{i,j\} \in E$. When the vertices are the causal states and the edges are the transition probabilities between causal states, the graph $G_{\epsilon}(\mathcal{S},\mathcal{T})$ is a special class of HMM called an \textbf{\eM{}}~\cite{CrutchPRL1989, epsilonMachines2}. The \eM{} both represents the memory and models future behaviour of a non-Markovian time series (Figure~\ref{fig:workflowHMM}z). For the On/Off duration sequences of quantum dots, the causal states have chemical and physical explanations such as the correlations obtained by having multiple brightness states, or varying emitting rates. We examine these examples and more in Section~\ref{sec:Discussion}.

We infer the causal states, \eMs{}, and complexities $C_{\mu}$ of the state duration sequence data $\mathbf{w}$ by using the \emph{Causal State Splitting Reconstruction} (\textbf{CSSR}) algorithm~\cite{CSSR2, CSSRbactra, CSSRmatlab}. The \eM{} output by CSSR is the model will represent non-Markovian component for the hypothesis testing in our study. We will refer to the model as the \emph{\eM{} Model} (\textbf{$\epsilon$MM}) hypothesis.

While we refer details of the algorithm to Ref.~\cite{CSSR2} and Appendix~\ref{apx:CSSR}, the successful operation of CSSR relies on some key attributes related to the input data. The first, is the algorithm requires as input the maximum length $L_{\text{max}}$ of past sequences to consider for the grouping of past observations in Equation~\eqref{eq:equivRelation}. If the non-Markovian memory contained within the time series has an intrinsic Markov order $L$, choosing $L_{\text{max}} < L$ results in poor prediction due to the inferred $\epsilon$MM not capturing the long-memory correlations in the data. However, CSSR will still produce a model that is consistent with the non-Markovian correlations up to order $L$. Given sufficient data, choosing $L_{\text{max}} \geq L$ guarantees convergence on the true extent of non-Markovian correlations. 

The second, is that given the finite length of data available, there is an upper limit on values of $L_{\text{max}}$ above which the resulting conditional probabilities $P(\vec{r} \;|\; \cev{r}_L)$ estimated in Equation~\eqref{eq:equivRelation} will be prone to severe under-sampling. This results in the algorithm creating new causal states for every string of length $L_{\text{max}}$ it encounters. To avoid this issue, the range of permissible past sequence lengths $L_{\text{max}}$ given data of length $N$ is bounded by 
\begin{gather}
    L_{\text{max}} \leq \frac{\log N}{\log |\mathcal{A}|},
    \label{eq:lmaxBound}
\end{gather}
where $|\mathcal{A}|$ are the number of unique measurement outcomes in the time series data~\cite{CSSR2,MartonAoP}. Generally speaking, the length of data $N$ suitable for CSSR is entirely dependent on the alphabet size and maximum desired Markov order. 

For the memoryless or Markovian hypothesis, the corresponding model can be deduced in simple cases. This can be done to anticipate the memory structure of the time series of alternating QD state durations in Equation~\eqref{eq:durationTimeseries}, should a blinking quantum dot store no information about its past behaviour. This is true if On/Off state durations $w$ are drawn independently from their respective distributions $P_{b}(w^b)$. The probability of observing the immediate future state duration $w^b_k$, would only need to know which brightness state $b$ the state duration at the previous timestep $w^b_{k-1}$ belonged to. With longer past sequences having no bearing on future behaviour, a blinking quantum dot with no correlations has an intrinsic Markov order 1.

The minimal memory model consists of two causal states $|\mathcal{S}| = 2$, one storing all past sequences with the most recent state durations coming from the On state $\cev{w} = (\ldots, w^{\text{Off}}_{k-2},w^{\text{On}}_{k-1})$, and the other for past sequences ending in a state duration from the Off state $\cev{w}' = (\ldots, w^{\text{On}}_{k-2},w^{\text{Off}}_{k-1})$. Any \eM{} that alternates between two causal states always has a memory of $C_{\mu} = 1$, and is shown in Figure~\ref{fig:workflowHMM}x. We refer to the alternating model with $C_{\mu} = 1$ as the \emph{alternating Markov model} (\textbf{AMM}), which can be constructed from alternating data $\v{w}$ by estimating the empirical distributions $\hat{P}_{\text{On}}(w) = \nu(w)/N_{\text{On}}$ and $\hat{P}_{\text{Off}}(w) = \nu(w)/N_{\text{Off}}$ from time series. The $\nu(w)$ term refers to the frequency counts for the outcome $w$ in the data, and $N_{\text{b}} : b\in\{\text{On},\text{Off}\}$ are the total number of On/Off symbols in the data respectively. The probabilities from $\hat{P}_{\text{On}}(w) = \nu(w)/N_{\text{On}}$ and $\hat{P}_{\text{Off}}(w) = \nu(w)/N_{\text{Off}}$ were used as the transition probabilities for the alternating HMM, arranged following the structure in Figure~\ref{fig:workflowHMM}x. We note that although an AMM and $\epsilon$MM when inferred at $L_{\max}=1$ may be the same, this is not always true in general~\cite{CSSR2}.

\subsection{Hypothesis testing}
\label{sec:HypTesting}

\begin{figure*}[t]
    \centering
    \includegraphics[width=0.7\textwidth]{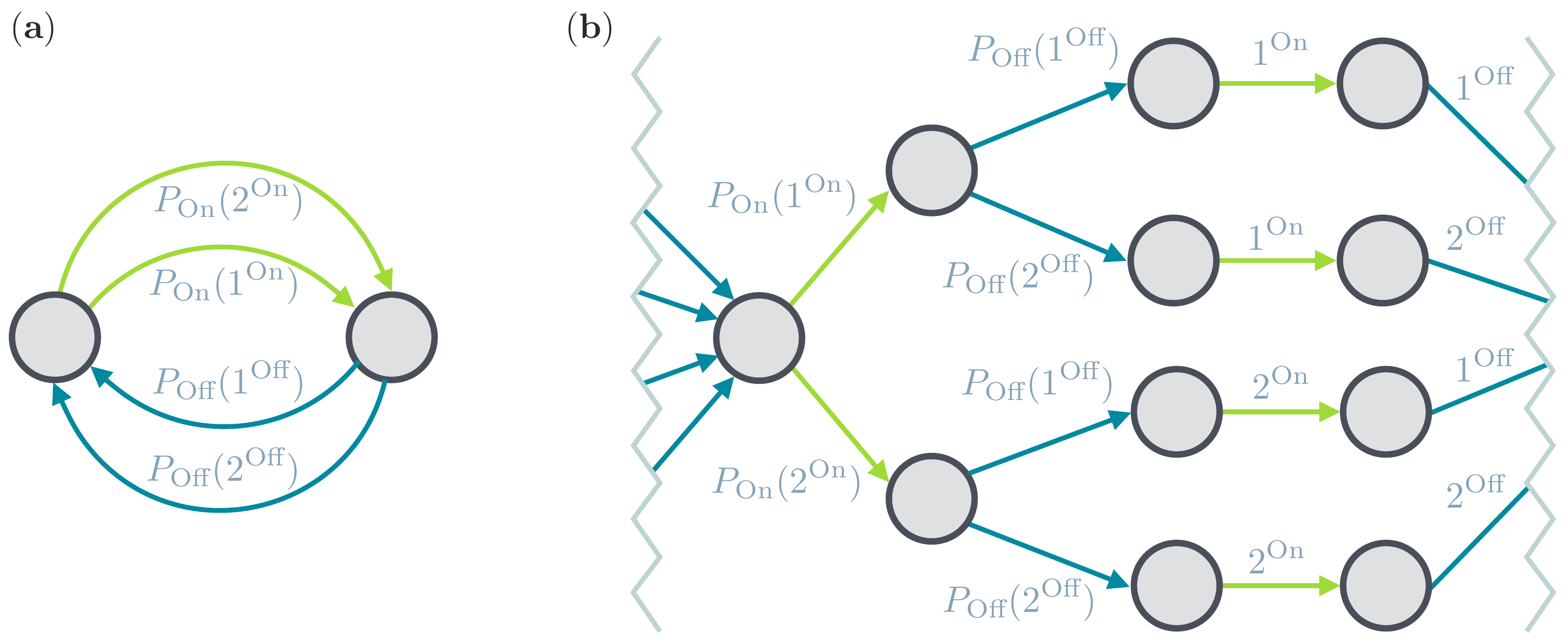}
    \caption{Examples of two HMMs that have different memory and which result in the same overall state duration distribution $P_b(w^b)$. The HMMs can be ``read" as follows: Pick a starting causal state (grey node), and follow any outgoing arrow to another causal state. This represents a measurement outcome which occurs with the labelled probability. Green arrows represent On measurement outcomes, blue arrows represent Off outcomes. If a causal state has only one outgoing arrow, then the labelled measurement outcome occurs with probability 1. Repeat these steps from the next causal state and so on, recording outcomes at each transition. \textbf{(a)} Example of an alternating Markov model (AMM) with two possible On and two possible Off durations, 1 and 2. \textbf{(b)} Example of a HMM with the same possible On/Off outcomes and probabilities as the AMM, but past sequences of length 2 are cloned. Starting from the leftmost state, an On duration followed by an Off duration are sampled from the distributions $P_{\text{On}}(w^{\text{On}})$ and $P_{\text{Off}}(w^{\text{Off}})$. The following two transitions clone the outcomes and then return back to the leftmost state to begin the process again. The On/Off duration distributions result the same graph for both models despite the HMM on the right having higher memory. The discrete probability distributions $P_b(w^b)$ are left unspecified for a purpose. They could be power-law, exponential, or something else entirely, and the overall structure of the two HMMs would not change. This also highlights that the specific distribution of the state durations does not indicate the presence or lack of memory.}
    \label{fig:echoAMM}
\end{figure*}

\emph{Complexity}, $C_{\mu}$ is the memory of an underlying process, with large values associated with a highly non-Markovian process. It is worth noting that while two HMMs can be statistically distinct in the future behaviours they produce, they may share the same $C_{\mu}$ and/or produce the same distribution over outcomes Equation~\eqref{eq:powerlaw-general}. Figure~\ref{fig:echoAMM} demonstrates an example of the latter. This highlights the need for being able to statistically test models when given data. Conducting statistical tests that use data to disprove models requires data $\v{r}$, a null hypothesis $H_0$, and an alternate hypothesis $H_1$. Unlike some other statistical tests, our test does not have a built-in or obvious $H_0$; we must use a pair of constructed models to conduct a test. 

To show that quantum dots have nontrivial memory in the blinking durations, our null hypothesis $H_0$ is the model with trivial memory, which is the AMM from Section.~\ref{sec:HMMandCSSR}. While the number of possible HMMs to use as the alternate hypothesis is large, selecting the model generated by CSSR has the advantage that it represents the \emph{minimal ideal} predictor~\cite{CSSR2}. This removes the need for the hypothesis test to penalise for overfitted hypotheses with too many causal states. 
Given the alternating Markov model as the null hypothesis $H_{\text{AMM}}$ (Figure~\ref{fig:workflowHMM}y), a HMM inferred by CSSR which may have $C_{\mu} \geq 1$ as the alternative hypotheses $H_{\text{$\epsilon$MM}}$ (Figure~\ref{fig:workflowHMM}z), and a general time series $\v{r}$, the \emph{negative log-likelihood ratio} (\textbf{nLLR}) either rejects or fails to reject $H_{\text{AMM}}$ under the rule,
\begin{equation}
    \begin{split}
        \ell(\v{r}) < \Gamma &\ \rightarrow\  \text{reject } H_{\text{AMM}} \\
        \ell(\v{r}) \geq \Gamma &\ \rightarrow\  \text{fail to reject } H_{\text{AMM}},
    \end{split}
    \label{eq:nLLRt}
\end{equation}

where
\begin{gather}
    \ell(\v{r}) = -2\log \frac{P(\v{r} | H_{\text{$\epsilon$MM}})}{P(\v{r} | H_{\text{AMM}})}
    \label{eq:nLLR}
\end{gather}
is the negative log-likelihood ratio (Figure~\ref{fig:workflowHMM}g). The $\Gamma$ in Equation~\eqref{eq:nLLRt} is the rejection threshold based on our chosen p-value $\alpha = 0.01$ which characterises the type-I error (false-positive) probability (Figure~\ref{fig:workflowHMM}g). It is given by
\begin{gather}
    P(\ell(\v{r}) \leq \Gamma \;|\; H_{\text{AMM}} \;\text{true}) = \alpha.
\end{gather}

Taken together, Equation~\eqref{eq:nLLRt} reads: when $\ell(\v{r}) < \Gamma$, then the AMM hypothesis is rejected by the data $\v{r}$ with significance $\alpha$; when $\ell(\v{r}) \geq \Gamma$ then the test fails to reject the AMM hypothesis (Figure~\ref{fig:workflowHMM}h). The probability of observing a sequence $\v{r}$ given a HMM is efficiently obtained via the forward algorithm~\cite{Rabiner1989,rabiner1986introduction}. Hence, we were able to compute the negative log-likelihood ratios with HMM hypotheses.

Unlike complexity or On/Off duration distributions, this statistical test can reject the minimal memory model. Rejection only occurs if sufficient data from a non-Markovian process is available (see Appendix~\ref{apx:dataQuality}).

\subsection{Non-Markovian memory in brightness state durations}
\label{sec:CGandResults}
Each time series of state durations (Equation~\eqref{eq:durationTimeseries}) consisted of a sequence of state duration times corresponding to each of the 7 quantum dots. Since different QD excitation powers induce different physical (and statistical) behaviour~\cite{Yuan2018}, we consider data from the \SI{100}{\nano\watt}\;-\;\SI{200}{\nano\watt} range which is expected to behave consistently (see Sec.~\ref{sec:Cand-3} for further details). Due to the wide range in the sizes of duration times for blinking quantum dots, the amount of unique measurement outcomes $|\mathcal{A}|$ for each time series was on the order of thousands. As mentioned in Sec.~\ref{sec:HMMandCSSR}, large $|\mathcal{A}|$ limits the capacity of CSSR to infer correlations over long Markov orders~\cite{CSSR2}. To combat this, each time series $\v{w}$ was coarse grained over measurement outcomes (Figure~\ref{fig:workflowHMM}f). We performed this by assigning each state duration $w^b$ appearing in a data sequence to a coarse grain symbol $\mathcal{C} : w^b \rightarrow c^b_x$ if $w^b \in \{[10^x,10^{x+1}) : x = 0,1,2,\ldots\}$. By using this coarse graining method, each coarse grain symbol $c_x^b$ corresponded to exponentially increasing intervals of brightness state durations:
\begin{equation}
    \begin{split}
        c_0^b &= [1,9] \\
        c_1^b &= [10,99] \\
        c_2^b &= [100,999] \\
        c_3^b &= [1000,99999],
    \end{split}
    \label{eq:compressionRule}
\end{equation}
for $b \in \{\text{On},\text{Off}\}$. Due to the power-law distributed behaviour of the state durations $P_b(w^b) \propto w^{-m_b}$, the coarse symbols shared similar behaviour, with ``short" duration times $c_0^b$ appearing most often, and ``long" \& ``extra-long" durations $c_2^b$ \& $c_3^b$ rare occurrences. While we chose to use 10 as the exponent base for the coarse interval sizes, \emph{post-hoc} tests showed that the main results of this study were unchanged when we used a moderately larger alphabet. The result of this procedure was a time series of alternating coarse On/Off state durations:
\begin{gather}
    \v{c} = (c^{\text{On}}_{x,1}, c^{\text{Off}}_{x,2}, \ldots, c^{b}_{x,k}, \ldots),
    \label{eq:compressTimeseries}
\end{gather}
shown in Figure~\ref{fig:workflowHMM}f.
The length of our data $\v{c}$ had a range of $N = [428,1080]$, with a post-compression alphabet size between $|\mathcal{A}| = [6,8]$. Using Equation~\eqref{eq:lmaxBound} the longest searchable memory length was $L_{\max}=3$. The $\epsilon$MMs inferred from the QD data in this study are valid as we do not violate this bound.

We constructed the alternating Markov model for each time series of coarse duration times, following the procedure in Sec.~\ref{sec:HMMandCSSR}. The \eMs{} for each time series data $\v{c}$ were inferred with CSSR over a range of Markov orders $L_{\text{max}}$, and the non-Markovian memory $C_{\mu}$ recorded. Finally, we performed nLLR tests between the AMM and $\epsilon$MM models for each coarse time series, and recorded the proportion of times the test rejected the alternating model.

Table~\ref{tab:mainResults} shows the non-Markovian memory inferred from the data as a function of probed Markov order $L_{\text{max}}$. When $L_\text{max}=1$, both models have trivial memory and thus the outcome of the statistical test is not relevant to our investigation. We observe that when $L_{\max} > 1$, the average non-Markovian memory contained in the coarse state durations of blinking quantum dots is greater than the minimal $C_{\mu} = 1$ hypothesis, within a spread indicated by the standard deviation $\text{std}(C_{\mu})$. Overall, the average non-Markovian memory $\langle C_{\mu} \rangle$ tended to increase with larger sequence lengths, which could imply that most quantum dots have nontrivial memory. In addition to the detection of non-Markovian memory in the data, the hypothesis testing revealed that the AMM is not sufficient to explain the majority of the statistical behaviour of the blinking. By assuming the minimal memory via AMMs, the measured data is significantly unlikely to occur. Therefore, we do not believe all the quantum dots have minimal memory.

Because processes which involve memory can be characterised by it, identifying the underlying drivers or conditions of the process can become more distinct when longer Markov orders are taken into account~\cite{Munoz2020}. Similarly, the ability for our methods to rule out memory hypotheses can be improved. Experimentally this is achieved by recording longer data. In our study, we were unable to analyse Markov orders greater than $L_{\max}=4$ due to the limitations in the length of measured data available (Equation~\eqref{eq:lmaxBound}), however the methods presented here are scalable for such future work. At high Markov orders, the distinction between memory-inducing chemical or physical effects could be made clearer. Diagnosing which of these effects are present in a QD may be achieved by finding theoretical bounds on the amount of memory they are allowed to induce, and comparing those to observations.

The main results presented in Table~\ref{tab:mainResults} only considered laser excitation powers between \SI{100}{\nano\watt} - \SI{200}{\nano\watt} for consistent physical behaviour~\cite{gibson2018excitation}. Higher excitation powers of \SI{400}{\nano\watt} - \SI{600}{\nano\watt} by contrast, are expected to behave differently. When the laser irradiance is higher, the quantum dot is more likely to capture two photons from the same laser pulse. As a result, sufficient energy is available to overcome the activation energy associated with erasing the memory. The expected effect is less memory in the state durations. We do indeed observe this effect for these excitation powers. At \SI{500}{\nano\watt} and \SI{600}{\nano\watt}, the AMM hypothesis was never rejected for all probed Markov orders (6 data sets @ \SI{600}{\nano\watt}, 14 data sets @ \SI{500}{\nano\watt}). Similarly, $11/23$ of the overall data exhibited memory and rejected the AMM for \SI{400}{\nano\watt}.

\begin{table}[t]
\centering
\begin{tabular}{|c|cccc|c|}
\hline
Markov order $L_{\max}$ & 1 & 2 & 3 & 4 & \multicolumn{1}{l|}{Laser Power} \\ \cline{1-6} \noalign{\vskip\doublerulesep\vskip-\arrayrulewidth} \cline{1-6}
Proportion reject AMM & 0/5 & 5/5 & 5/5 &  & \multirow{3}{*}{\SI{100}{\nano\watt}}  \\ 
$\langle C_{\mu} \rangle$ & 1.000 & 1.875 & 1.908 &  & \\
$\text{std}(C_{\mu})$ & 0.00 & 0.43 & 0.40 & & \\ \cline{1-6}
Proportion reject AMM & 0/5 & 4/5 & 4/5 & 2/2 & \multirow{3}{*}{\SI{200}{\nano\watt}}\\ 
$\langle C_{\mu} \rangle$ & 1.000 & 1.573 & 1.566 & 2.347 &  \\ 
$\text{std}(C_{\mu})$ & 0.00 & 0.33 & 0.33 & 0.27 & \\ \hline
\end{tabular}
\caption{Results of non-Markovian memory analysis and hypothesis testing of models for blinking quantum dot data as a function of Markov order $L_{\max}$. For each time series of coarse-grained On/Off state durations $\v{c}$, the proportion of time series which rejected the AMM model via the nLLR test, average non-Markovian memory $\langle C_{\mu}\rangle$, and standard deviation of memory $\text{std}(C_{\mu})$, is shown.}
\label{tab:mainResults}
\end{table}

\section{Explaining non-Markovianity in blinking}
\label{sec:Discussion}

To justify the presence of non-Markovian memory in the blinking quantum dot data, we simulate several hypothetical blinking processes and test them.  We determine if they explain the observed non-Marokvianity in Table.~\ref{tab:mainResults}. We show physical or chemical process cause a model which has a simple AMM structure, to gain non-Markovian memory. For each memory candidate presented in this section, we simulate its effect on a toy model and show when it strays significantly from the alternating, $C_{\mu} = 1$ hypothesis. We perform hypothesis testing for each candidate source of memory based on the negative log-likelihood ratio method outlined in Sec.~\ref{sec:HypTesting}. In this section we also rule out systematic data processing treatments we applied as a source of non-Markovianity. All simulations referred to in this section were performed using self-written MATLAB code.

\subsection{Poisson toy model}
\label{sec:PoissonModel}
\begin{figure}[t]
    \centering
    \includegraphics[width=\columnwidth]
    {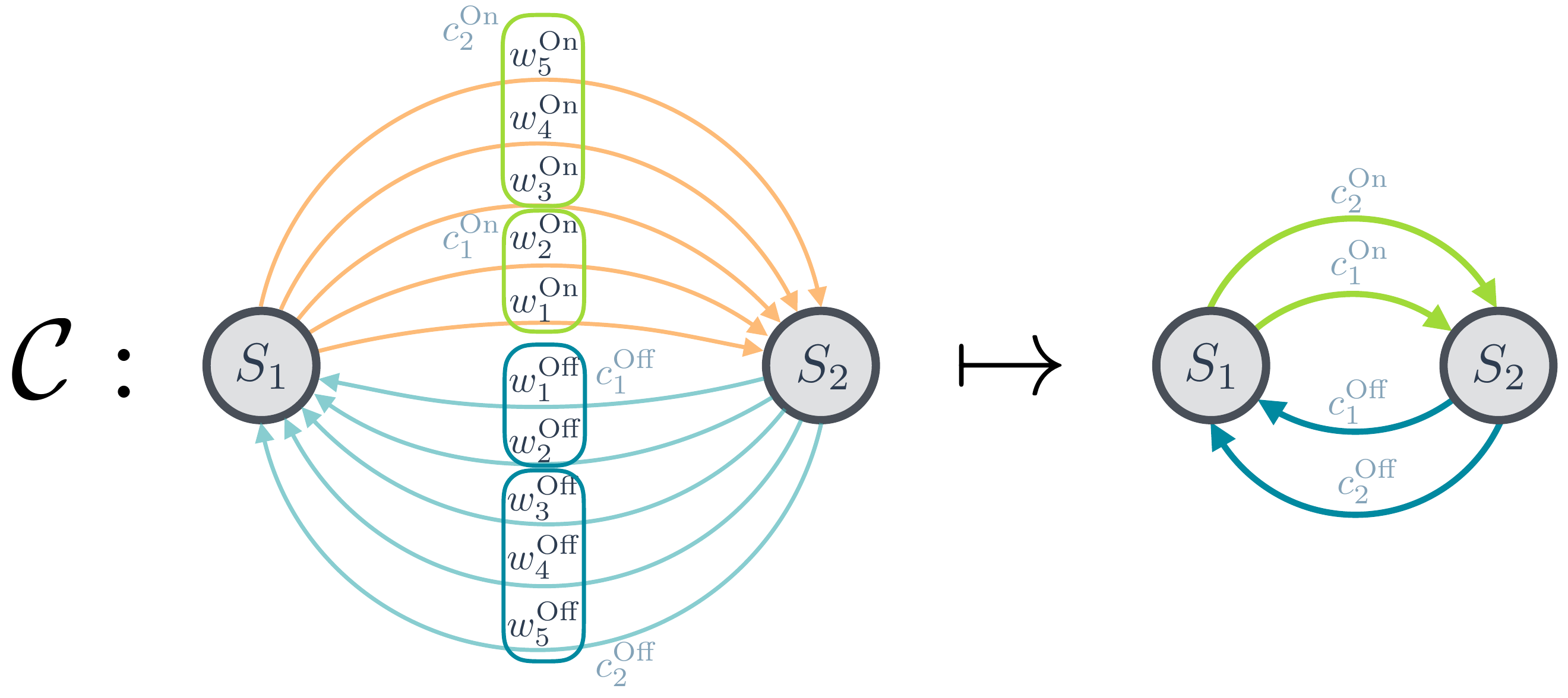}
    \caption{Example of the effect of a coarse mapping $\mathcal{C}$ on the alternating Markov model. The AMM on the left is at the state duration level. Gold and blue transition probabilities correspond to the ``On" and ``Off" state duration times respectively. Coarse graining groups together On/Off durations $w^b$ into coarse divisions $c^b$, represented by the coloured boxes, and results in the AMM on the right. The number and distribution of causal states $S_i$ in the AMM before and after coarse graining is the same, and $C_{\mu} = 1$ in both cases. Hence, so long as the coarse graining only groups together common On and Off states, the Markovian property is preserved.}
    \label{fig:coarseWait}
\end{figure}

For a non-blinking quantum dot emitting photons at a constant rate $\lambda_{\text{On}} > 0$, the probability of observing $n$ photons arriving in the detector over an observation duration $T$ is Poisson distributed:
\begin{align}
    P_{\text{On}}(n) = \frac{(\lambda_{\text{On}} T)^n}{n!} e^{-\lambda_{\text{On}}T},
    \label{eq:poissOn}
\end{align}
where the $n$ arrival times are distributed uniformly and randomly over $T$. An \eM{} which simulates this process would consist of a single causal state with one self-transition probability corresponding to $P(\text{Arrival}) = \lambda_{\text{On}}$ and another corresponding to $P(\text{No arrival}) = 1-\lambda_{\text{On}}$. Having only one causal state, a single Poisson process acting alone in this manner is an example of a memoryless process~\cite{marzen2015renewal, marzen2017renewal}. For the toy model to simulate a blinking emitter, a Poisson process with a background rate $ \lambda_{\text{Off}}$ corresponding to the dark state is introduced. A chosen master observation duration $T$ is then divided into sub-intervals of size $w$. The sizes of the sub-intervals are power-law distributed
\begin{gather}
    P_b(w^b) = \frac{(w^b)^{-m_b}}{H_{w^b_*}^{(m_b)}} \;:\; b\in\{\text{On},\text{Off}\},
    \label{eq:powerlaw-wait}
\end{gather}
where $m_b > 1$ is the power-law constant, and $w^b_*$ is a chosen upper limit on the size of the longest possible sub-interval. The constant $H_{w^b_*}^{(m_b)}$ is the generalised harmonic number and is related to the normalisation of the distribution. By choosing unique power-law constants $m_{\text{On}}$ and $m_{\text{Off}}$ in addition to maximum sub-interval lengths $w_*^{\text{On}}$ and $w_*^{\text{Off}}$ for the bright and dark states, we sample the sequence $w_1^{\text{On}}, w_2^{\text{Off}}, \ldots, w_T$ until the master interval is filled. The sequence of alternating sub-intervals are exactly the state duration time series $\v{w}$ mentioned in Sec.~\ref{sec:powerlaws}. The arrival times of a blinking quantum dot are finally simulated by substituting $w_k^b \rightarrow T$ into Equation~\eqref{eq:poissOn}, sampling Poisson arrival times for each waiting time sub-interval in the sequence. Since each Poisson distribution is sampled in an alternating way, the blinking toy model is an example of  the alternating HMM described in Sec.~\ref{sec:HMMandCSSR} and shown in Figure~\ref{fig:coarseWait} and has $C_{\mu} = 1$ over the waiting times as desired. We note that choice of waiting time distribution doesn't change the memory of an alternating HMM at the waiting time level. So long as the waiting times between the On/Off states are independently distributed we are free to choose whichever distribution we like. We chose the waiting times to be power-law distributed to match the general behaviour of the On/Off duration statistics of the experimental data. This procedure results in simulated arrival times of a memoryless blinking quantum dot with highly tuneable statistical behaviour. The goal of constructing the simulated blinking quantum dot in this way is that we are able to rule out systematic data processing as a source of the observed non-Markovianity. Recalling the data processing outlined in the aforementioned sections, the two main treatments we applied, were the binning of the arrival times to intensities (Figure~\ref{fig:workflow}c), and the coarse graining of the waiting times (Figure~\ref{fig:workflowHMM}e-f). 

The first is ruled out via the properties of a Poisson process: for an emitter where photons arrive at a constant rate $\lambda_{\text{On}}$, the probability of observing $n$ photons over \emph{any} time window is Poisson distributed as in  Equation~\eqref{eq:poissOn}. Binning the nanosecond-resolution arrival times into millisecond windows $\Delta t$ is equivalent to replacing $T$ with $\Delta t$ in Equation~\eqref{eq:poissOn}, and thus the statistics of the emitter remain Poisson distributed independent of the size of the binning window. Since Poisson processes are inherently memoryless, this implies they also remain so independently of the choice bin size in counting arrival times. 

Figure~\ref{fig:coarseWait} illustrates that coarse graining the emission symbols of an alternating process does not increase memory. Starting from the $C_{\mu}=1$ alternating HMM, the possible waiting times for the bright state are represented by gold arrows stemming from a single causal state $S_1$. Similarly, all possible waiting times for the dark state are shown as blue arrows stemming from another causal state $S_2$. Since all the On/Off waiting times each stem from their own respective causal state, this signals that waiting times for an intensity level $\{w^b\}$ are statistically independent and identically distributed under some distribution $P_{\text{On}}(w)$ and $P_{\text{Off}}(w)$, having no unique future behaviour conditional of which specific On/Off waiting time was observed. By applying any coarse graining $\mathcal{C}$ over the waiting times in an alternating HMM, the effect is to group transition probabilities stemming from the same state together. The new HMM possesses the same transition structure as the original alternating HMM, where the coarse waiting times in Equation~\eqref{eq:compressionRule} remain statistically independent, and now identically distributed under some other distribution $P'_{\text{On}}(c^{\text{On}})$ and $P'_{\text{Off}}(c^{\text{Off}})$. 
In both cases, the number and distribution of the causal states is invariant before and after coarse graining, and hence the non-Markovian memory given by Equation~\eqref{eq:Cmu} is not affected. The demonstration of memory invariance under both aforementioned transformations secures the observed memory as stemming from the quantum dot's behaviour and/or properties, rather than our methodology.

\subsection{Detector errors}
\label{sec:DetectorErrors}
The optics, avalanche photodiode, and clock that collect photon arrival times may have errors which we refer to as ``detector errors". To rule out these errors as a cause of memory, we simulate two cases where such errors might be present and examine how a blinking $C_{\mu}=1$ emitter may be affected by them. The first is the simple case of background and dark counts, where the photodiode detects a photon arrival when it shouldn't. The second case is the converse \emph{false-off} scenario, where the detector fails to register a photon when it should. This is possible in a number of ways. Firstly, large portions of photons are not detected due to the limited numerical aperture of the optical system. Some photons are not detected owing to the quantum efficiency of the avalanche diode.  Finally, photons may not be detected if they arrive within the dead time of the avalanche diode or the clock, but this is a typically negligible cause. 

Assuming that background and false-off detector errors are memoryless and occur independently of the emitting sample~\cite{buchinger1995}, their effect on the memory in the waiting times is the respective addition and removal of arrivals at a constant rate $\lambda_{\text{err}}$. For a blinking emitter with photon arrival rates $\lambda_{\text{On}}$ and $\lambda_{\text{Off}}$, memoryless and independent detector errors will always manifest as a shift in the emitting rates of the brightness states as $\lambda_{\text{On/Off}}' = \lambda_{\text{On/Off}} \pm \lambda_{\text{err}}$. 
Here the result is another memoryless blinking emitter with Poisson photon arrival rates $\lambda_{\text{On/Off}}'$.

The sum of background photons and dark counts are measured by removing the emitting sample. Adjusting the On/Off threshold method (Figure~\ref{fig:workflow}c-d,u) to not classify any measured intensity given under that condition as part of the bright state, characterises this error. While the false-off error rate in a given experiment involves the emitting Poisson rate $\lambda_{\text{On}}$, the optical aperture, avalanche photodiode efficiency, and efficiency of the optics used, the false-off rates can generally be neglected by choosing quantum dots that display sufficiently distinguishable brightness states. There are common cases where false-off rates in a detector could be high enough for the bright peak of an intensity histogram to be shifted down to almost background levels. While this type of data is usually discarded, this makes the distinction between the brightness states of the QD unclear which in turn has the potential to shift memory away from the alternating $C_{\mu} =1$ model. Having obtained data that had a clear separation between the bright and dark states, we find it unrealistic to expect this amount of classification fault our study.

\subsection{Classification}
\label{sec:Classification}
The lack of distinction between brightness states ultimately does influence the amount of non-Markovian memory one can observe, and thus we examine which other processes leading to low contrast of brightness states could explain memory found in the blinking patterns. The following subsections deal with potential sources of non-Markovian memory which all have the effect of introducing an ambiguity in brightness state classification during the thresholding step. We simulate four such cases related to plausible physical or chemical processes which may occur in a general blinking quantum dot and examine their influence on statistical complexity under the change of a key parameter. For each candidate, we inferred an ensemble of $\epsilon$MM from the simulated data and performed nLLR tests to find whether the memory structure deviated significantly from the AMM hypothesis. 

\begin{figure*}[ht]
    \centering
    \includegraphics[width=0.9\textwidth]{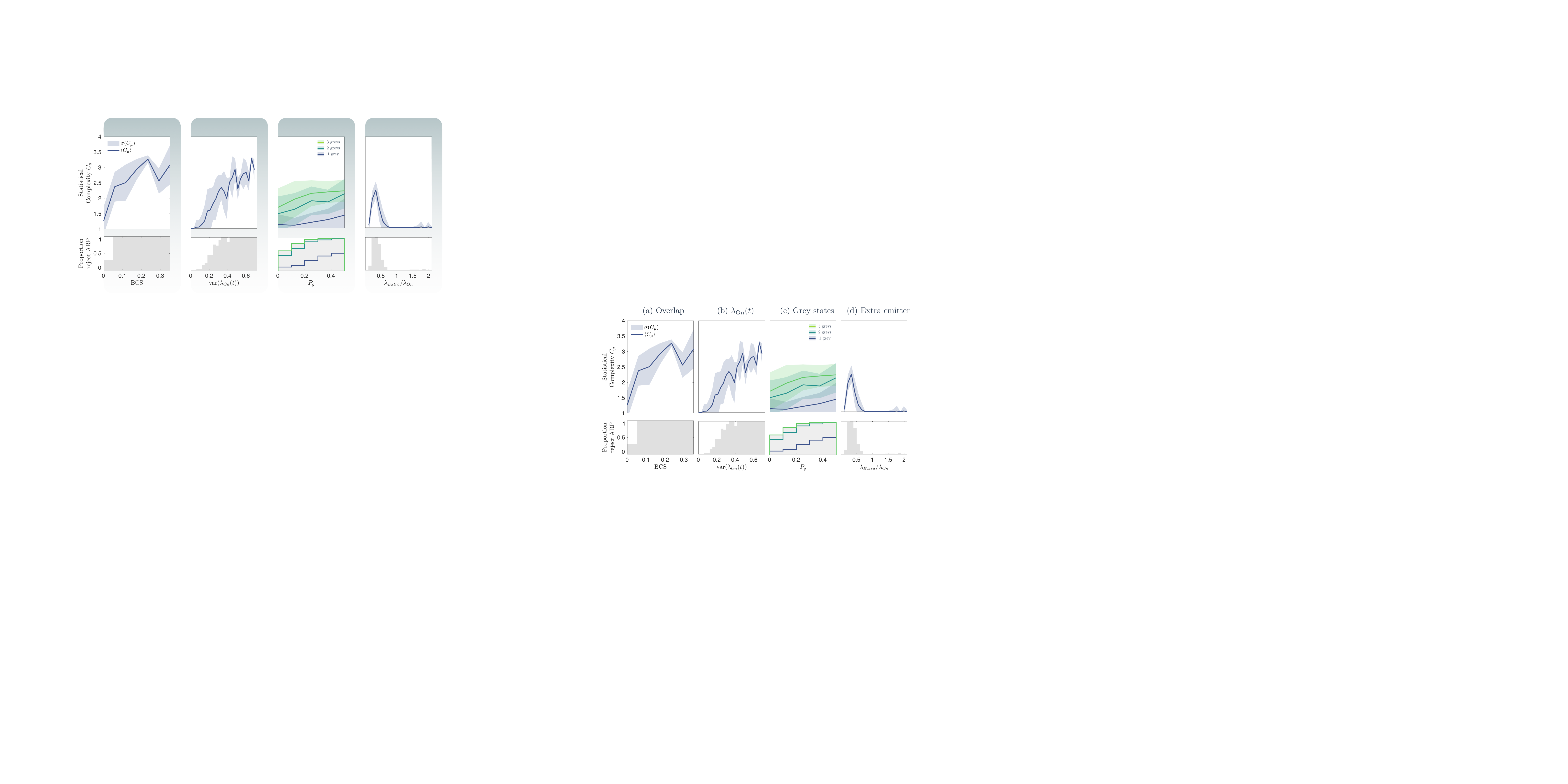}
    \caption{Memory and hypothesis testing results for each memory candidate. Top rows show mean memory $\langle C_{\mu} \rangle$ and the standard deviation $\sigma (C_{\mu})$. Bottom row shows the proportion of simulations which rejected the AMM hypothesis. The scale of the y-axes in each row is consistent.
    \textbf{(a)} Memory and hypothesis testing as a function of On/Off peak overlap quantified by BCS in Equation~\eqref{eq:BCS-closed}.
    \textbf{(b)} Memory and hypothesis testing as a function of $\lambda_{\text{On}}$ fluctuation strength, quantified by the variance var$(\lambda_{\text{On}}(t))$. All simulations with high fluctuation strengths var$(\lambda_{\text{On}}(t)) > 30$ exhibit memory $C_{\mu} > 1$.
    \textbf{(c)} Memory and hypothesis testing vs. probability $P_g$ of transitioning to any of the $N_g$ grey states. Each ribbon shows the mean memory and standard deviations for a blinking emitter with 1, 2, and 3 grey states. Each solid line in the hypothesis testing histogram shows the proportion of simulations which significantly rejected the AMM hypothesis, given a number of grey states. 
    \textbf{(d)} Memory and hypothesis testing vs. the brightness of an extra emitter $\lambda_{\text{Extra}}$ relative to a reference $\lambda_{\text{On}} = 2.5$ arrivals/10 time steps.
    }
    \label{fig:allCands}
\end{figure*}

\subsubsection{Low state contrast}
\label{sec:Cand-1}
\textbf{Model:} Since we classify brightness states takes by choosing a threshold between peaks on a PL intensity histogram, the simplest way in which brightness state ambiguity may be introduced is in a process where $\lambda_{\text{On}}$ after whatever error, is sufficiently close to $\lambda_{\text{Off}}$. Overall the effect is realised as an amount of overlap between PL intensity peaks, which themselves are probability distributions. The PL intensity histogram peaks are approximately Poisson distributed. This is excluding minor effects which distort the statistics such as excess PL intensity due to the QD switching brightness states during the binning time window $\Delta t$. Assuming exactly Poisson statistics allow us to construct a similarity rating which describes the overlap between the PL intensity peaks, and quantifies how indistinguishable the bright and dark states are. While many information-geometric quantities which all relate to a distance between probability distributions exist~\cite{amari2016book}, the Bhattacharyya coefficient (\textbf{BC})~\cite{bhattacharyya1946,amari2016book} is a well-suited starting point for describing overlap. Given two discrete distributions $\mathcal{P}$ and $\mathcal{Q}$ over some domain $\mathcal{X}$, the BC describes the amount of overlap as the intersected probability between them
\begin{gather}
    \text{BC}(\mathcal{P},\mathcal{Q}) = \sum_{x \in \mathcal{X}} \sqrt{p(x)q(x)},
    \label{eq:Bhattacharyya}
\end{gather}
where $0 \leq BC \leq 1$, and $BC = 1$ implies the distributions are fully-overlapped and hence, identical. For the case of the brightness peaks on a PL intensity histogram, the full intensity distribution $\mathcal{R}$ is normalised instead of the component bright and dark peaks. For the proportion of intensity events coming from the bright distribution $n_{\text{On}}$, the proportion of events coming from the dark distribution is $1-n_{\text{On}}$. The full PL intensity distribution is then described by a convex combination between the bright and dark Poisson curves ($\mathcal{P}$ and $\mathcal{Q}$ respectively),
\begin{align}
    \mathcal{R} &\sim n_{\text{On}}\mathcal{P} + (1-n_{\text{On}})\mathcal{Q}.
    \label{eq:convexDistribution}
\end{align}

Referring to the components of the convex mixture as $\mathcal{P}' = n_{\text{On}}\mathcal{P}$ and $\mathcal{Q}' = (1-n_{\text{On}})\mathcal{Q}$, we define the amount of overlap between them as
\begin{gather}
    \text{BCS}(\mathcal{P}', \mathcal{Q}') \coloneqq 2\sqrt{n_{\text{On}} (1- n_{\text{On}})}\text{BC}(\mathcal{P},\mathcal{Q}).
    \label{eq:BCS}
\end{gather}
Constructing similarity between the bight and dark states in this manner has some nice properties, and interpretation. Effectively it is the regular overlap between the distributions normally expected by the Bhattacharyya coefficient, scaled by how balanced the relative heights the bright and dark Poisson peaks are in the full intensity distribution $\mathcal{R}$. The factor of 2 is included such that $0 \leq \text{BCS} \leq 1$, where $\text{BCS} = 1$ implies maximum overlap like for the BC in Equation~\eqref{eq:Bhattacharyya}. Maximum similarity between brightness states is achieved when each state is mixed equally $n_{\text{On}} = 1/2$, and when $\lambda_{\text{On}} = \lambda_{\text{Off}}$. For Poisson-distributed $\mathcal{P} \sim \text{Pois}(\lambda_{\text{On}}\Delta t)$ and $\mathcal{Q} \sim \text{Pois}(\lambda_{\text{Off}}\Delta t)$, a closed-form expression~\cite{calin2014} is given by
\begin{gather}
    \text{BCS}(\mathcal{P}',\mathcal{Q}') = 2\sqrt{n_{\text{On}} (1- n_{\text{On}})} \;e^{\left(-\frac{1}{2}(\lambda_{\text{On}}\Delta t - \lambda_{\text{Off}}\Delta t)^2\right)}
    \label{eq:BCS-closed}
\end{gather}
which we ultimately use to quantify brightness state overlap in our simulations. The $\Delta t$ term in Equation~\eqref{eq:BCS-closed} accounts for the freedom of different choices of arrival time binning windows (10ms in our case). 

\textbf{Simulation:} The statistical complexity versus $\lambda_\text{On}$ was found by simulating photon arrival times of a blinking QD over a master interval of $T = 500,000$ time steps, and subdividing it into power-law sized sub-intervals of waiting times with constants $m_{\text{On}} = 1.2$ and $m_{\text{Off}} = 1.1$. The relative heights between the bright and dark intensity peaks were modulated by allowing the maximum waiting time of a single bright state sub-interval $w_*^{\text{On}}$ to iterate between $2\%$ and $10\%$ of the master interval $T$. Background rates were fixed to $\lambda_{\text{Off}} = 0.1$ arrivals/time-step, with the amount of overlap modulated by varying the bright arrival rate between $0.6 \leq \lambda_{\text{On}} \leq 2.5$ arrivals/time-step. To obtain intensity distributions for the each sequence of simulated arrival times, arrivals were binned over $\Delta t = 10$ time step windows. Where possible, we defined On/Off thresholds using the same methods outlined in Sec.~\ref{sec:dataDesc} for the data. The BCS for each iteration of $w_*^{\text{On}}$ and $\lambda_{\text{On}}$ was calculated. Post thresholding, the simulated data was converted to waiting times which were then coarse grained using the exponentially-scaled bins of base 10 mentioned in Sec.~\ref{sec:CGandResults}.  The \eMs{} and statistical complexities $C_{\mu}$ for the simulations were inferred by passing the coarse-grained state duration time series into CSSR at the longest Markov order $L_{\max}$ available. The relationship between BCS and $C_{\mu}$ is plotted in Figure~\ref{fig:allCands}a, in addition to the proportion of simulations with memory structure significantly different to the $C_{\mu} = 1$ alternating hypothesis as determined by nLLR testing.

\textbf{Results:} Figure~\ref{fig:allCands}a shows that additional memory above an AMM increases as similarity increases. With a increase in On/Off state similarity, the likelihood of observing long waiting times post-thresholding decreases drastically due to the proliferation of short-length classification errors. Since the area between the On/Off intensity peaks corresponds to intensities that have been misclassified, this means long sequences of intensities originally stemming from one brightness state are likely to be interrupted by short misclassifications. For the AMM model shown in Figure~\ref{fig:coarseWait}, the probability of observing a short state duration at timestep $k$ was only dependent on which brightness state $b$ a past sequence ended in. With overlap, sequences ending in long durations $\cev{c} = (\ldots, c^{b'}_{x,k-2}, c^b_{\text{long},k-1})$ have higher probabilities of producing a short duration in the next time step. Since observing a long duration in the last time step $c^b_{\text{long},k-1}$ causes statistically different future behaviour compared to seeing any other duration, new causal states and memory are needed to account for this.

We also observe that all simulations significantly reject the alternating model for $\text{BCS}\geq 0.05$. This suggests that very low overlap is required before the simulated quantum dots can no longer be described by the alternating model. It is likely for the data used in this study to have greater overlap than this. Although additional memory simply due to low brightness contrast is a possible effect in our data, it is one that can be tested for in an any general experiment. This is performed by simply observing good separation between PL intensity peaks, or filtering out quantum dots with low brightness contrast.

\subsubsection{Time evolving $\lambda(t)$}
\label{sec:Cand-2}
\textbf{Model:} Each brightness state of the quantum dot consists of at least two Hamiltonian eigenstates: an excited state and a ground state. It is possible that there are actually many eigenstates with similar, but not quite identical, brightnesses. The thresholding procedure classifies all these eigenstates as the same brightness state. The result allows the emitting rate $\lambda_{\text{On}}$ of the quantum dot to fluctuate randomly in time. Thermal expansion of the optical bench can also cause random changes in the instrument detection efficiency, contributing to this effect. 

\textbf{Simulation:} As before, we simulated arrival times by subdividing a master time interval of $T = 500,000$ time steps into power-law sized sub-intervals with constants $m_{\text{On}} = 1.2$ and $m_{\text{Off}} = 1.1$, and sampling On/Off arrival times within them. We set the single largest possible waiting times $w_*^b$ for the bright and dark states to $8\%$ and $1\%$ of the master interval $T$ respectively. We chose these values of $w_*^b$ following preliminary testing which found them to be suitable for producing a wide range of simulations where On/Off states could be classified from the intensity distribution. Arrival times were binned over $\Delta t = 10$ time step windows.  We simulated a fluctuating bright rate by setting $\lambda_{\text{On}}$ to an initial rate of $\lambda_{\text{On}}(0) = 2.5$ arrivals/time-step and allowing it to evolve as a random walk between the interval $[0.3, 3.0]$ in steps of $0.1$ for each bright state sub-interval. The same effect could be achieved by allowing $\lambda_{\text{On}}$ to fluctuate with every time step in the data, not just during bright sub-intervals. In practice however, locating during which brightness state allows the emitting rate to fluctuates in a blinking QD is nontrivial, hence method suffices as a simplified model. For each iteration of the random walk lower bound, we ran $50$ independent simulations. Background rates were fixed to $\lambda_{\text{Off}} = 0.1$. An automatic thresholding script was used to define the On/Off states following the classification convention in Sec.~\ref{sec:dataDesc} from the PL intensity data where possible, after which the simulations were converted to waiting time sequences and coarse-grained using the method in Sec.~\ref{sec:CGandResults}. The $\epsilon$MMs and statistical complexities $C_{\mu}$ were inferred from each simulation by sending the coarse-grained time series into CSSR, searching at the largest possible Markov order $L_{\max}$. Electing to keep the random walk step size fixed, we used the lower bound of the random walk as the key parameter which modulated the strength of the fluctuations. This is because decreasing the lower bound allowed for a higher chance for the variance of the emitting rate fluctuations to increase. By setting the lower bound of the random walk close to the background rate $\lambda_{\text{Off}}$, the simulated effect could be achieved of the random walk of $\lambda_{\text{On}}$ tending mostly downward. 

\textbf{Results:} Using the variance of the emitting rate random walk Var $\lambda_{\text{On}}(t)$ as the summary of fluctuation strength, the response of statistical complexity $C_{\mu}$ is shown in Figure~\ref{fig:allCands}b. While an increase in additional memory is possible across most fluctuation strengths, we observe that this is rare when the fluctuations from the initial emitting rate $\lambda_{\text{On}}(0)$ are small. This is evident by the strong grouping of simulations resulting in AMM structures for Var $\lambda_{\text{On}}(t) < 0.20$. Above this level, the proportion of simulations with an inferred memory structure significantly different than the alternating $C_{\mu} = 1$ hypothesis, rises above $50\%$ until it becomes mostly saturated at Var $\lambda_{\text{On}}(t) \geq 0.35$.
Taken together, the results in Figure~\ref{fig:allCands}b demonstrate that even when the chemical processes which drive the fluctuations of emitting rates are random, an amount of non-Markovian memory between the lifetimes of brightness levels is incurred. There is no \emph{a priori} reason to expect the random walk of $\lambda_{\text{On}}$ as implemented here is exactly realised in the sense that changes occur only at the beginning of bright states. However, if the emitting rate is allowed to fluctuate freely in time, the extent of the fluctuations may be estimated. This could be done by performing change point analysis (CPA)~\cite{palstra2021python,watkins2005detection,ensign2009bayesian,ensign2010bayesian} on the intensity data. The number of change points occurring within a band of high intensity, would give an indication of multiple bright eigenstates.

\subsubsection{Grey states}
\label{sec:Cand-3}
\textbf{Model:} Classifying a time series of PL intensity measurements into bright and dark states is common practice in studies of blinking emitters. While we separated the PL intensity of the QDs in this study into two brightness states, there may be in general states of intermediate brightness present in quantum dot photoluminescence. The intermediate brightness states, referred to as \emph{grey states}~~\cite{Spinicelli2009}. Grey states occur when a quantum dot undergoes a reversible change in chemical configuration~\cite{gomez2009exciton}, resulting in a lower, but nonzero photoluminescence quantum yield. The number of chemical configurations available is presumably large, but the number of observed quantum yield values is small. Grey states are thought to be present in a variety of systems, and can be identified as excess counts, or a bump between the bright and dark peaks in a PL intensity histogram (See Figure~3b in Ref.~\cite{Yuan2018}). Grey states are observed for the QDs in this study at high laser powers (\SI{300}{\nano\watt} - \SI{600}{\nano\watt}). Although the main analysis techniques in our study used lower laser powers which excluded QDs with prominent grey states (Table~\ref{tab:mainResults}), our methodology still generalises to handle the non-Markovianity which may arise from them. Therefore, for this demonstration we set the thresholding procedure to consistently include grey states as part of the dark state. The alternating $C_{\mu}=1$ model shown in Figure~\ref{fig:coarseWait} assumes only two intensity levels, bright and dark, where a set of coarse-grained waiting times is assigned to each (e.g. $\mathcal{W}_{\text{On}} = \{$short, medium, long$\}$ and $\mathcal{W}_{\text{Off}} = \{$short, medium, long$\}$ waiting times for both bright and dark intensity levels respectively). One might consider adding an extra set of alphabet letters for each additional grey state $\mathcal{W}_{g}$. However, larger alphabets limit the length of non-Markovian memory that can be inferred via CSSR (see Sec.~\ref{sec:HMMandCSSR}). Instead, the lifetime dynamics of the grey states must be embedded within the bright and dark states. This would be likely to involve a memory cost for doing so which we test for here.

\textbf{Simulation: }To simulate a blinking emitter with a number of possible grey states, we followed the same procedure outlined in Sec.~\ref{sec:PoissonModel}, subdividing a master observation window $T = 500,000$ time steps into power-law sized waiting times with constants $m_{\text{On}} = 1.2$ and $m_{\text{Off}} = 1.1$. We allowed the single largest possible waiting times for the bright and dark states to be fixed to $8\%$ and $1\%$ of the master interval $T$ respectively. However, instead of sampling photon arrival times alternating between the bright and dark states, we included grey states and followed a simple rule for determining how to transition between intensity levels. Among the QDs in this study that do exhibit grey states, only 1 intermediate level is present~\cite{Yuan2018}. However, other systems such as perovskite quantum dots demonstrate evidence of multiple grey states~\cite{park2015room,li2018excitons,seth2016fluorescence}, and thus we generalise to that scenario here. Given a specific number of desired grey states $N_g$, we let $P_g$ be the probability of transitioning to any one grey state from the bright or dark states. For each sub-interval waiting time, the corresponding intensity level was determined via the following $(2+N_g) \times (2+N_g)$ transition matrix:
\begin{gather}
\v{T}_{\stackbin{\text{\scriptsize{bright}}}{\text{\footnotesize{state}}}} =\! 
    \kbordermatrix{
    & \text{On} & \text{Off} & \text{Grey 1} & \cdots & \text{Grey }N_g\\
    \text{On} & 0 & \! 1-P_g \! & \frac{P_g}{N_g} & \cdots & \frac{P_g}{N_g} \\
    \text{Off} & 1-P_g & 0 & \frac{P_g}{N_g} & \cdots & \frac{P_g}{N_g} \\
    \stackbin{\text{\scriptsize{Grey}}}{1} & \frac{1}{2} & \frac{1}{2} & 0 & \cdots & 0 \\
    \vdots & \vdots & \vdots & \vdots & \ddots & \vdots \\
    \stackbin{\text{\scriptsize{Grey}}}{N_g}
     \! & \frac{1}{2} & \frac{1}{2} & 0 & \cdots & 0
    }.
    \label{eq:tmatrixGrey}
\end{gather}
In this example, the simulated quantum dot has a probability of $1/2$ to transition to either bright or dark state from any grey state, and for simplicity we deny the possibility for a grey state to transition to another grey state. When $P_g$ is small, the dominant dynamics are alternating between the bright and dark states, with the intermediate grey state transitions acting as noise on the PL intensity levels. We simulated up to $N_g = 3$ grey states in this example. Each state used power-law sized sub-intervals with constants $m_g = 1.2$ and allowed the largest state lifetime to be $2\%$ of the master interval $T$. For the arrival times, we fixed the background and emitting rates to be $\lambda_{\text{Off}} = 0.1$ and $\lambda_{\text{On}} = 2.5$. Photon arrival rates for each grey state were evenly spaced between the interval $[\lambda_{\text{On}}$, $\lambda_{\text{Off}}]$. Arrival times were binned over $\Delta t = 10$ time step windows. The probability $P_g$ of entering into any grey state was varied between $0.05$ and $0.5$. For each iteration we ran $50$ independent simulations. Each simulation of intensity data was then sent through a thresholding algorithm, coarse-grained, and passed into CSSR at the longest Markov order $L_{\max}$ available. The response of $C_{\mu}$ with respect to $P_g$, in addition to the nLLR testing is shown in Figure~\ref{fig:allCands}c. 

\textbf{Results: }On average, non-Markovian memory increases with the overall probability $P_g$ of entering into a grey state. The increase is shown to be proportional to the number of additional grey states $N_g$ simulated in this example. Both of these effects can be attributed due to the rise in intermediate intensity levels as $P_g$ and $N_g$ increases. It should be noted that while on average memory is increased above $C_{\mu} = 1$, there are still some instances where the alternating model is \emph{accepted} via nLLR hypothesis testing compared to CSSR. The lower subplot of Figure~\ref{fig:allCands}c demonstrates this. For $N_g = 1$ the proportion of times AMM is rejected is never greater than $50\%$, despite high grey transition probability and a mean complexity $\langle C_{\mu} \rangle> 1$. This effect is reflected in a fraction of the CdSe quantum dots used in this study. At high excitation powers, nearly all quantum dots demonstrated evidence of grey states yet were better described by the AMMs. For multiple grey states however $N_g = 2$ or 3, this is not the case. The proportion of simulations with memory structure significantly different than the alternating model rises and saturates quickly with $P_g$. 
Identifying intermediate intensity levels in a quantum dot is nontrivial when the pathways leading to their emissions are rare. In such instances, identifying their presence via PL intensity histograms alone is difficult and one would turn to more sophisticated methods~\cite{palstra2021python,Yuan2018,yuan2017degradation}.

\subsubsection{Multiple emitters}
\label{sec:Cand-4}
\textbf{Model:} The final case we consider is when two independent quantum dots are mistaken as a single emitter. This occurs when two quantum dots are chemically bonded together or, by chance, are so close together that an Abbe diffraction-limited microscope cannot distinguish them~\cite{rust2006,nehme2018,whelan2015super}.
This effect was simulated by taking two blinking emitters with rates $\lambda_{\text{On}}$ and $\lambda_{\text{Extra}}$, sampling arrival times independently from each of them, and them superimposing the arrivals. The first emitter was taken as the ``reference" with $\lambda_{\text{On}} = 2.5$. The second extra emitting rate $\lambda_{\text{Extra}}$ was varied by iterating its value as multiples of $\lambda_{\text{On}}$, ranging from $0.1\lambda_{\text{On}}$ to $2.0\lambda_{\text{On}}$. The range of the extra emitter has a dual interpretation. If $\lambda_{\text{Extra}} = \lambda_{\text{On}}$, then the simulation mimics two quantum dots with the same quantum yield. Otherwise it mimics two quantum dots with different quantum yields.

\textbf{Simulation: }For each iteration of $\lambda_{\text{Extra}}$, 25 independent simulations were run. Background rates for were fixed to $\lambda_{\text{Off}}$. The single longest sub-interval waiting times $w_*^b$ for the reference and extra emitter were set to $8\%$ and $0.6\%$ of the master interval $T = 500,000$ time steps. For the background, this was set to the usual $1\%$. Power law constants for the size of the sub-intervals were $m_{\text{On}} = m_{\text{Extra}} = 1.2$, and $m_{\text{Off}} = 1.1$. After superimposing the reference and extra emitter's arrival times, the data was binned over $\Delta t = 10$ time step windows to obtain the PL intensity sequence. These sequences where split into On/Off categories and then converted into a sequence of coarse-grained state lifetimes. The memory structure was inferred from the simulations via CSSR, and nLLR tests were performed for each.

\textbf{Results:} Figure~\ref{fig:allCands}d shows the response of statistical complexity $C_{\mu}$ against the ratio $\lambda_{\text{Extra}}/\lambda_{\text{On}}$, in addition to the nLLR hypothesis testing.
Aside from minor statistical fluctuations for $\lambda_{\text{Extra}}/\lambda_{\text{On}} \geq 1.4$, the memory obtained by introducing a second QD is observed to not increase above $C_{\mu}=1$ for a large range of relative difference in emitting rates. A high-memory region between $0.2 \leq \lambda_{\text{Extra}}/\lambda_{\text{On}} \leq 0.4$ is identified where the majority of simulations reject the alternating model. At this proportion of relative emitting rates, the extra QD has a rate positioned in between the background $\lambda_{\text{Off}}$ and reference $\lambda_{\text{On}}$ rates. This leads to a similar scenario mentioned in Sec.~\ref{sec:Cand-1} where the excess of intensity counts between the On/Off peaks causes correlations between the state lifetimes, due to classification ambiguity by the threshold procedure. However, the photon antibunching and excitation intensity dependence~\cite{Yuan2018} are distinct from grey state phenomena.

\section{Conclusion}
\label{sec:Conclusion}
We have shown that memory is observed in the sequences of On/Off state durations by using tools from computational mechanics, and hypothesis testing. By modelling the memory as hidden Markov Models, we have devised a hypothesis testing method and ruled out the AMM model which corresponds to trivial memory in the data. For 90\% of the \SI{100}{\nano\watt} \& \SI{200}{\nano\watt} QD data used in this study, the statistical test rejected the AMM model when Markov orders $L_{\max}>1$ were taken into account (Table~\ref{tab:mainResults}). Rejecting the AMM model indicates the existence of long-range correlations in the state duration data. We highlight that the duration histograms used in characterising blinking behaviour are unable to reveal non-Markovianity, and we show that observing them to be exponential or power-law distributed gives no information regarding this (Figure~\ref{fig:echoAMM}).

By pairing our methods with a robust simulation, we were able to simultaneously rule out or associate memory in QD simulations to a number of specific chemical, physical, and experimental effects. Specifically, low state contrast, fluctuating emitting rates $\lambda_{\text{On}}$, and the presence of grey states were all associated with an increase of memory for simulated data. In relation to the experimental data, grey states were not present in the \SI{100}{\nano\watt} and \SI{200}{\nano\watt} range, and hence not a cause for the observed memory. Although further experimental analysis outside the scope of this paper is required in order to determine the relevancy of On/Off contrast (Equation~\eqref{eq:BCS-closed}) and fluctuating $\lambda_{\text{On}}$ as cause for memory in our data, we have discussed ways to identify them from QD blinking data in their relevant sections. We also ruled out time binning, detector errors, and extra identical emitters as effects associated with memory, in both the simulations and experimental data.

It is tempting to assume the aforementioned effects associated with memory may be explained simply due to ambiguity in On/Off state classification, observed via an excess in events between On and Off peaks (Figure~\ref{fig:workflow}u)~\cite{gopich2006theory,gopich2005theory,yip1998classifying}. However, we discussed using the grey state example in Section~\ref{sec:Cand-3} why this is not so. This affirms the notion that the observation of memory in both data and simulation involves more than just any process which results in low On/Off state contrast. This counters the argument that memory in blinking QD data can be solely inferred from the intensity histograms, and strengthens the foundation of our analysis.

 Although we considered specific examples in Section~\ref{sec:Classification}, our methods are general enough to be applied to other chemical systems. One way to test for other intrinsic memory processes would be to perform this analysis on QD data over a range of compositions and measurement techniques. For example, observing memory in systems with blinking rates different to those seen in our data~\cite{ghosh2021}, would be an indicator of other chemical processes which involve memory that we have yet to identify. Furthermore, that information could be used to predictably determine the conditions and effects required for unblinking dots in general. The results for the QDs in this study excited at high laser powers (Section~\ref{sec:CGandResults}) present an example where experimental conditions can be associated with non-Markovianity.

Overall, finding memory in QD blinking has allowed us to relate it to specific physical influences that can be experimentally verified. The examples considered in this study are by no means a complete list~\cite{gomez2009exciton}. Yet we are confident that the rejection of time binning (Figure~\ref{fig:workflow}c) and systematic false off or background errors introduced via the detector (Section~\ref{sec:DetectorErrors}) as sources of memory, ensures the robustness of our methods to be applied to a wide variety of data measured under similar assumptions. 

Ultimately what can be predicted is still an open question which will require more collaboration between chemistry and information theory. For quantum dots with memory, each $\epsilon$MM has a drastically different graph structure. This difference could be related to differences in quantum dot crystal or surface structure. Our methods do not determine that an $\epsilon$MM is a correct model of quantum dot memory.  It is possible that substantially larger datasets would reveal a better model, which might be a slight modification of the $\epsilon$MM found here, or a model with far more memory. However, the scope of potential questions that may be answered via our methods is large. This is made possible by the scalable design which can readily accept longer data, and requires only that the number and outcomes of measurements are finite. Even if this is not the case, there is a body of theoretical work covering HMMs and memory in such cases~\cite{marzen2017renewal,marzen2017structure,brodu2020discovering}, thus making a pathway for our procedure to be extended to more general data. In any case, the observation of memory in QD blinking and its relation to chemical processes in our study, serves as a first step towards discovering a richer and practical understanding of quantum dots. If we can identify a chemical synthesis method that causes quantum dots to `remember' a hidden Markov model that increases their time-averaged quantum yields, quantum dot lighting will become more efficient.

\appendix
\label{apx:Main}
\section{Details of CSSR algorithm}
\label{apx:CSSR}
The CSSR algorithm~\cite{CSSR2} proceeds to construct sets of causal states accounting for past sequences of outcomes, iterating over the length of the past sequences starting at $L_{\max} = 0$. In each iteration, the algorithm first estimates the probabilities $P(r_k|\cev{r}_{L})$ of observing a symbol conditional on each possible length $L$ past sequence and compares them with the distribution $P(r_k | \mathcal{S} = S_i)$ it would expect from the causal states it has so far reconstructed. If $P(r_k|\cev{r}_L) = P(r_k | \mathcal{S} = S_i)$ for some causal state, then the past sequence $\cev{r}_L$ is identified with it since it results in the same future behaviour as any other sequence in $S_i$. If the probability is found to be different for all existing $S_i$, then a new causal state is created to accommodate the future behaviour caused by observing the sub-sequence. By constructing new causal states only as necessary, the algorithm guarantees a minimal model that describes the non-Markovian behaviour of the data (up to a given Markov order), and hence the corresponding \eM{} of the process.

The CSSR algorithm compares probability distributions via the \emph{Kolmogorov-Smirnov} \textbf{(KS)} test~\cite{Massey1951, Hollander2013}. The hypothesis that $P(r_k|\cev{r}_L)$ and $P(r_k | \mathcal{S} = S_i)$ are identical up to statistical fluctuations is rejected by the KS test at the significance level $\sigma$ when a distance $\mathcal{D}_{KS}$ is greater than tabulated critical values of $\sigma$~\cite{Miller1956}. The KS distance is given by
\begin{gather}
    \mathcal{D}_{KS} = \sup_{r_k} | F(r_k | \mathcal{S} = S_i) - F(r_k | \unexpanded{\cev{r}}_L)|,
    \label{eq:KSdistance}
\end{gather}
where $F(r_k | \mathcal{S} = S_i)$ and $F(r_k | \unexpanded{\cev{r}}_L)$ are cumulative distributions of $P(r_k | \mathcal{S} = S_i)$ and $P(r_k | \unexpanded{\cev{r}}_L)$ respectively. The supremum $\sup_{r_k}$ means that the distance $\mathcal{D}_{KS}$ considers only the largest absolute difference between $F(r_k | \mathcal{S} = S_i)$ and $F(r_k | \unexpanded{\cev{r}}_L)$. The significance $\sigma$ sets a limit on the accuracy of the history grouping by parametrising the probability that an observed history $\cev{r}_L$ belonging to a causal state $S_i$, is mistakenly split off and placed in a new causal state $S_j$. Preliminary testing demonstrated the output of CSSR to be robust under small changes in $\sigma$. Our analysis used the default value $\sigma = 0.001$.

\section{Data quality for hypothesis testing}
\label{apx:dataQuality}

\begin{figure}[t]
    \centering
    \includegraphics[width=\columnwidth]{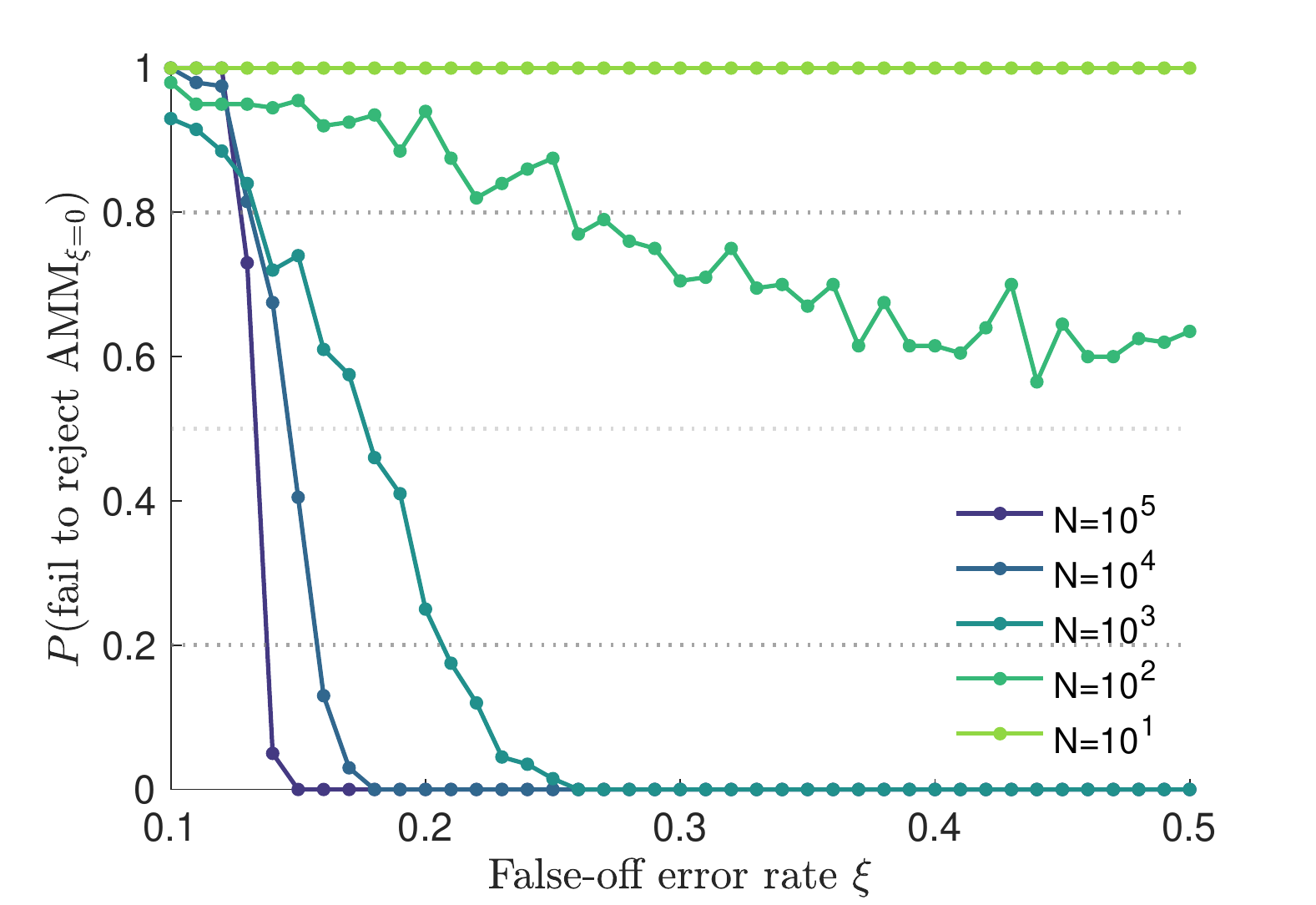}
    \caption{Probability of hypothesis test failing to reject AMM$_{\xi=0}$ as a function of false-off error probability $\xi$ in simulated AMM data. Hypothesis testing performed using AMM$_{\xi=0}$ as the null and AMM$_{\xi=0.5}$ as the alternate hypotheses.}
    \label{fig:falseOff}
\end{figure}

The reliability of the hypothesis testing methods in Section~\ref{sec:HypTesting} depends on the length and quality of data sequences available in an experiment. The reliability of a hypothesis test is often characterised by the probability for the test to reject the null when the alternate is true. By exploring this quantity as a function data length and data quality, we devised a scenario using simulated AMMs to provide an estimate on the data conditions needed to use the methods presented in this paper. 

This was done as follows: A sequence of AMM state durations $\v{w}$ following Equation~\eqref{eq:durationTimeseries} was simulated using an AMM (Figure~\ref{fig:workflowHMM}x). Copies of $\v{w}$ were made and false-off errors (Section~\ref{sec:DetectorErrors}) were introduced at constant probabilities between $\xi = [0.1,0.5]$. All sequences were compressed using Equation~\eqref{eq:compressionRule} to form sequences $\v{c}$ following Equation~\eqref{eq:compressTimeseries}. 

We fit two AMMs from the compressed sequences (Figure~\ref{fig:workflowHMM}y): one from the sequence with no error, and another from the sequence with 50\% false-off error. To distinguish them, we labelled them by the amount of false off error, e.g. AMM$_{\xi = 0.5}$. We defined AMM$_{\xi = 0}$ as the null hypothesis and AMM$_{\xi = 0.5}$ as the alternate hypothesis. We performed nLLR testing between them using the cloned sequences, truncating them at varying lengths $N$. Tests were performed at a significance level $\alpha = 0.01$.

The results shown in Figure~\ref{fig:falseOff} demonstrate that an increase in the length of data dramatically improves the ability of the test to reject the null when the data approaches being better described by the alternate model. While for the extreme cases of short data sequences $N=10^1$ \& $N=10^2$, the hypothesis test struggles to distinguish between the two example models, a length of $N=10^3$ appears sufficiently accurate. Ultimately the point at $\xi=0.5$ is when $H_1$ is true. Since sequences $N\geq 10^3$ always reject $H_0$ at this point, the statistical power of the test is very high in this example.

The point at which the curve goes below 0.5 on the y-axis is where the hypothesis test is able to identify perturbations in the data above chance. For increasing $N$, this occurs at smaller error rates, indicating that the increased sequence lengths increase the sensitivity of the hypothesis testing. For $N=10^5$ this happens at around $\xi \approx 0.15$ which is assumed to be far above what is generally present from an experiment. This implies that the hypothesis test can be reliably performed without modelling the specific error.

While testing between the two models mentioned here is a specific example, this demonstration serves as a heuristic example. In the framework of blinking data in general, we estimate sequence lengths $N \geq 10^3$ (i.e. $10^3$ blinks) and false-off error rates $\xi < 0.15$ are optimal for our methods. Both of these are achievable in the majority of blinking experiments.

\bibliography{references}

\end{document}